\documentclass[10pt,a4paper]{article}
\usepackage[left=2cm,right=2cm,top=2cm,bottom=2cm]{geometry}

\usepackage{amsmath}
\usepackage{hyperref}

\numberwithin{equation}{section}
\newcommand{\bea}{\begin{eqnarray}}
\newcommand{\eea}{\end{eqnarray}}
\newcommand{\be}{\begin{equation}}
\newcommand{\ee}{\end{equation}}
\newcommand{\nn}{\nonumber \\}

\begin{document}
\begin{titlepage}
\phantom{.}
\vskip 1.5cm
\begin{center}
{\bf\Large ${\cal N}=8$ supersymmetric mechanics with spin variables
\vspace{0.2cm}

from indecomposable multiplets}
\end{center}
\vspace{1cm}

\begin{center}
{\large\bf Evgeny Ivanov${\,}^{a)\,b)}$, Stepan Sidorov${\,}^{a)}$}
\end{center}
\vspace{0.4cm}

\centerline{${\,}^{a)}$ \it Bogoliubov Laboratory of Theoretical Physics, JINR, 141980 Dubna, Moscow Region, Russia}
\vspace{0.2cm}
\centerline{${\,}^{b)}$ \it  Moscow Institute of Physics and Technology,
141700 Dolgoprudny, Moscow Region, Russia}
\vspace{0.3cm}

\centerline{\tt eivanov@theor.jinr.ru, sidorovstepan88@gmail.com}
\vspace{0.2cm}

\vspace{2cm}

\par

{\abstract \noindent We define two new indecomposable (not fully reducible) ${\cal N}=8$, $d=1$ off-shell multiplets and consider the corresponding models of
${\cal N}=8$ supersymmetric mechanics with spin variables. Each multiplet is described off shell by a scalar superfield
which is a nonlinear deformation of the standard scalar superfield $X$ carrying the $d=1$ multiplet ${\bf (1,8,7)}$. Deformed systems involve, as invariant subsets, two different off-shell
versions of the irreducible multiplet ${\bf (8,8,0)}$.
For both  systems we present the manifestly ${\cal N}=8$ supersymmetric superfield constraints, as well as the component off- and on-shell invariant actions, which for one version
exactly match those given in \href{https://arxiv.org/abs/2402.00539v3}{arXiv:2402.00539 [hep-th]}.
The two models differ off shell, but prove to be equivalent to each other on shell, with the spin variables sitting in the adjoint representation of the maximal $R$-symmetry group ${\rm SO}(8)$. } \vfill{}


\end{titlepage}

\section{Introduction}
Supersymmetric mechanics with spin variables was proposed in \cite{FIL0812}
as a system based on the coupled dynamical and semi-dynamical (with the first order kinetic terms for physical bosonic $d=1$ fields) irreducible multiplets. Recently, a new approach
to such systems based on the not fully reducible (indecomposable) multiplets was proposed \cite{SS2410}
\footnote{Such $d=1$ supermultiplets are known under various names: ``non-minimal'' \cite{Toppan1006,Toppan1204}, ``tuned'' \cite{Hubsch1310}, ``long'' and ``indecomposable'' \cite{CasimirEnergy,IS1509}.
Their characteristic feature is the presence of some subset of fields forming an irreducible supermultiplet on their own.}.

The first example of ${\cal N}=8$ supersymmetric model with spin variables was presented in \cite{FI2402}, using an off-shell ${\cal N}=4$ superfield formalism. Later on, the authors of ref. \cite{KhKN2408} demonstrated that
this model on shell is  invariant  under ${\cal N}=8$, $d=1$ superconformal group OSp$(8|2)$. In the original paper \cite{FI2402}, it was conjectured that the model constructed corresponds to an indecomposable ${\cal N}=8$
supermultiplet encompassing the dynamical ${\bf (1,8,7)}$ multiplet with a pair of  semi-dynamical ${\bf (8,8,0)}$ multiplets. The basic aim of the present paper is to prove this conjecture. We define the relevant ${\cal N}=8$
superfield set subjected to the proper manifestly ${\cal N}=8$ supersymmetric nonlinear constraints and show that the off-shell field content considered in \cite{FI2402} naturally comes out as a solution of these
constraints. We also introduce another indecomposable ${\cal N}=8$ multiplet with the same number of physical bosonic and fermionic fields, construct the relevant  off- and on-shell actions and find that on shell it yields
the same system as in \cite{FI2402}.

The superfield description of the multiplet ${\bf (1,8,7)}$ was originally given within the ${\rm SU}(2)\times{\rm SU}(2)\times{\rm SU}(2)$ covariant formalism in \cite{ABC}.
The direct superfield construction of manifestly ${\cal N}=8$ invariant actions has not been accomplished as yet. The invariant actions were constructed in the framework
of harmonic ${\cal N}=4$, $d=1$ superspace \cite{ABC, DI0706, FI1810}.
The covariant description of the multiplet ${\bf (1,8,7)}$ with the maximally extended $R$-symmetry ${\rm SO}(7)$ was presented in \cite{Toppan0511},
at the component level. The invariant Lagrangian was constructed there by ``brute force'' method.

The multiplet ${\bf (8,8,0)}$ was studied in the ${\rm SO}(4)\times {\rm SO}(4)$ covariant formulation in \cite{ABC}.
When dealing with this multiplet in the present paper, we rely on ref. \cite{ILS1807}, where ${\cal N}=8$, $d=1$ superalgebra was deformed to the superalgebra $su(4|1)$.
Here we are interested only in the undeformed case, with the standard ${\cal N}=8$ Poincar\'e superalgebra written in the ${\rm SU}(4)$ covariant form.
As was shown in \cite{ILS1807},  ${\cal N}=8$  multiplets with the off-shell $d=1$ field content ${\bf (8,8,0)}$ can exist in three nonequivalent SU$(4)$ covariant forms
related to each other by triality which is the specific feature of the complete $R$-symmetry group ${\rm SO}(8)$ of ${\cal N}=8$ supersymmetry.
We studied only two of such multiplets since the third one requires a different $d=1$ superspace for its off-shell description.
Here, we also ignore the third version for the same reason.
The triality  property refers also to the multiplets with the content ${\bf (0,8,8)}$. On the other hand, SU$(4)$ covariant formulations of the multiplets
${\bf (1,8,7)}$, ${\bf (2,8,6)}$, ${\bf (6,8,2)}$ and ${\bf (7,8,1)}$ are unique.

The paper is organized as follows. In Section \ref{m187} we start with the irreducible multiplet ${\bf (1,8,7)}$.
First, we present its SU(4) covariant description and derive invariant off-shell component Lagrangians through the proper reductions of the Lagrangians of the $ {\cal N}=8$ multiplet ${\bf (2,8,6)}$.
We also recall the superfield construction in terms
of harmonic ${\cal N}=4$ superfields accommodating the multiplets ${\bf (1,4,3)}$ and ${\bf (0,4,4)}$.
Further, we rewrite the constraints of the multiplet ${\bf (1,8,7)}$
in the maximally SO(7) covariant form. In Section \ref{v1},
we describe the first version of the indecomposable ${\cal N}=8$ multiplet unifying the multiplet ${\bf (1,8,7)}$
with a pair of ${\bf (8,8,0)}$ multiplets. We deform the constraints of the multiplet ${\bf (1,8,7)}$
by adding complex constrained superfields describing these two ${\bf (8,8,0)}$ multiplets.
We construct the component off-shell Lagrangian and show its identity with that of the model constructed  in \cite{FI2402}.
In Section \ref{v2}, we consider a different version of the indecomposable ${\cal N}=8$ multiplet,
in which the ${\bf (1,8,7)}$ multiplet is coupled to the pair of the second multiplet ${\bf (8,8,0)}$ from the list given in  \cite{ILS1807}.
In this version, the superfield constraints of the multiplet ${\bf (1,8,7)}$ are deformed by chiral and tensor $ {\cal N}=8$ superfields.
We construct its off-shell Lagrangian. We also compare the on-shell Lagrangians
of both indecomposable multiplets and prove their equivalence.
In Section \ref{con}, we summarize the results and discuss some problems for the future analysis.


\section{Multiplet ${\bf (1,8,7)}$}\label{m187}

\subsection{SU(4) covariant formulation}
We deal with the ${\cal N}=8$, $d=1$ superalgebra written as
\bea
    \left\lbrace Q^{I}, \bar{Q}_{J}\right\rbrace = 2\,\delta^{I}_{J}H,\qquad
    \left[H, Q^{I}\right]=0,\qquad
    \left[H, \bar{Q}_{I}\right]=0.\label{algebra_su41}
\eea
Here, the capital indices $I, J$ ($I=1,2,3,4$) refer to the ${\rm SU}(4)$ fundamental representation.
The corresponding superspace consists of a time coordinate $t$ and complex Grassmann coordinates $\theta_I$\,, $\overline{\left(\theta_I\right)}=\bar{\theta}^I$.
Supersymmetry acts on them as
\bea
    \delta\theta_{I} = \epsilon_{I}\,,\qquad
    \delta\bar{\theta}^{I} = \bar{\epsilon}^{I},\qquad
    \delta t = i\left(\bar{\epsilon}^{I}\theta_{I} + \epsilon_{I}\bar{\theta}^{I}\right),\qquad \overline{\left(\epsilon_I\right)}=\bar{\epsilon}^I.\label{SS}
\eea
One can easily define left and right time coordinates:
\bea
    t_{\rm L} =t+i\,\bar{\theta}^I\theta_I\,,\qquad t_{\rm R} =t-i\,\bar{\theta}^I\theta_I\,.
\eea
The left chiral subspace transforms as
\bea
    \delta\theta_{I}=\epsilon_{I}\,,\qquad \delta t_{\rm L}=2i\,\bar{\epsilon}^I\theta_I\,.\label{left_tr}
\eea
The covariant derivatives are
\bea
    D^I=\frac{\partial}{\partial\theta_I} - i\,\bar{\theta}^I\partial_t\,,\qquad
    \bar{D}_I=-\,\frac{\partial}{\partial\bar{\theta}^I}+i\,\theta_I \partial_t\,.\label{covDSU4}
\eea
This worldline realization follows from the realization of \cite{ILS1807} in the flat-superspace limit.

An unconstrained scalar superfield $X$ has ${\bf 256}$ components in its $\theta$-expansion.
The number of components is reduced to ${\bf 16}$ by imposing the following quadratic constraints\footnote{The antisymmetric tensor $\varepsilon^{IJKL}$ satisfies the following identities:
\bea
    &&\varepsilon^{1234}=1,\qquad \varepsilon_{1234}=1,\qquad \varepsilon^{IJKL}\varepsilon_{IJKL}=24,\nn
    &&\varepsilon^{IJKL}\varepsilon_{IJKM}=6\,\delta^{L}_{M}\,,\qquad \varepsilon^{IJKL}\varepsilon_{IJMN}=4\,\delta^{K}_{[M}\,\delta^{L}_{N]}\,,\qquad \varepsilon^{IJKL}\varepsilon_{IMNP}=6\,\delta^{J}_{[M}\,\delta^{K}_{N}\,\delta^{L}_{P]}\,.\nonumber
\eea}
\bea
    D^I\bar{D}_J X - \frac{\delta^I_J}{4}\,D^K\bar{D}_K X = 0,\qquad
    D^I D^J X - \frac{1}{2}\,\varepsilon^{IJKL}\,\bar{D}_K\bar{D}_L X=0,\label{187}
\eea
The superfield $X$ is expanded as
\bea
    X\left(t,\theta,\bar{\theta}\right) &=& x + \theta_{I}\psi^{I} - \bar{\theta}^{I}\bar{\psi}_{I}  - \frac{i}{2}\left(\theta_{I}\theta_{J}+\frac{1}{2}\,\varepsilon_{IJKL}\,\bar{\theta}^K\bar{\theta}^L\right)A^{IJ}+ \bar{\theta}^I\theta_I\,C\nn
    &&+\,i\,\bar{\theta}^I\theta_I\left(\theta_{J}\dot{\psi}^{J} + \bar{\theta}^{J}\dot{\bar{\psi}}_{J}\right)-\frac{i}{3}\,\varepsilon^{IJKL}\,\theta_{I}\theta_{J}\theta_{K}\dot{\bar{\psi}}_{L}+\frac{i}{3}\,\varepsilon_{IJKL}\,\bar{\theta}^I\bar{\theta}^J\bar{\theta}^K\dot{\psi}^{L}\nn
    &&+\,2\left[\left(\theta\right)^4+\left(\bar{\theta}\,\right)^4\right]\ddot{x}-\frac{1}{2}\left(\bar{\theta}^I\theta_I\right)^2\ddot{x}-2i\left[\left(\theta\right)^4-\left(\bar{\theta}\,\right)^4\right]\dot{C}+\frac{1}{2}\,\bar{\theta}^{K}\theta_{K}\theta_{I}\theta_{J}\dot{A}^{IJ}+\frac{1}{2}\,\theta_{K}\bar{\theta}^{K}\bar{\theta}^I\bar{\theta}^J\dot{A}_{IJ}\nn
    &&
    -\,2\left(\theta\right)^4\bar{\theta}^I\ddot{\bar{\psi}}_{I}+2\left(\bar{\theta}\,\right)^4\theta_I\ddot{\psi}^{I}-\frac{1}{2}\left(\bar{\theta}^I\theta_I\right)^2\left(\theta_{J}\ddot{\psi}^{J} - \bar{\theta}^{J}\ddot{\bar{\psi}}_{J}\right)\nn
    &&+\,\frac{1}{6}\left(\bar{\theta}^I\theta_I\right)^3\ddot{C}- \frac{i}{2}\left[\left(\theta\right)^4+\left(\bar{\theta}\,\right)^4\right]\left(\theta_{I}\theta_{J}+\frac{1}{2}\,\varepsilon_{IJKL}\,\bar{\theta}^K\bar{\theta}^L\right)\ddot{A}^{IJ}\nn
    &&+\,\frac{i}{2}\left[\left(\theta\right)^4\varepsilon_{IJKL}\,\bar{\theta}^I\bar{\theta}^K\bar{\theta}^L\left(\partial_t\right)^3 \psi^{J} - \left(\bar{\theta}\,\right)^4\varepsilon^{IJKL}\,\theta_{I}\theta_{J}\theta_{K}\left(\partial_t\right)^3 \bar{\psi}_{L}\right]+\left(\theta\right)^4\left(\bar{\theta}\,\right)^4\left(\partial_t\right)^4 x,\nn
    &&\overline{\left(x\right)}=x,\quad\overline{\left(\psi^{I}\right)}=\bar{\psi}_{I}\,,\quad A^{IJ}=-\,A^{JI},\quad
    \overline{\left(A^{IJ}\right)}=\frac{1}{2}\,\varepsilon_{IJKL}\,A^{KL}=A_{IJ}\,,\quad
    \overline{\left(C\right)}=C,\label{X}
\eea
where
\bea
    \left(\theta\right)^4 = \frac{1}{24}\,\varepsilon^{IJKL}\,\theta_{I}\theta_{J}\theta_{K}\theta_{L}\,,\qquad \left(\bar{\theta}\,\right)^4 = \frac{1}{24}\,\varepsilon_{IJKL}\,\bar{\theta}^{I}\bar{\theta}^{J}\bar{\theta}^{K}\bar{\theta}^{L}.
\eea
So this superfield describes an irreducible off-shell multiplet with the content ${\bf (1,8,7)}$.
The constraints \eqref{187} respect the $R$-symmetry group SO$(7)\subset {\rm SO(8)}$,
though only its ${\rm SU}(4)\sim{\rm SO}(6)$ subgroup is manifest. The ${\cal N}=8$ supersymmetry transformations of the component fields read:
\bea
    &&\delta x = \bar{\epsilon}^{I}\bar{\psi}_{I}-\epsilon_{I}\psi^{I},\qquad\nn
    &&\delta \psi^{I} = i\,\epsilon_{J}A^{IJ}+\bar{\epsilon}^{I}\left(i\dot{x}+C\right),\qquad
    \delta \bar{\psi}_{I} = -\,i\,\bar{\epsilon}^{J}A_{IJ}-\epsilon_{I}\left(i\dot{x}-C\right),\nn
    &&\delta C = -\,i\left(\bar{\epsilon}^{I}\dot{\bar{\psi}}_{I}+\epsilon_{I}\dot{\psi}^{I}\right),\qquad \delta A^{IJ} = 4\,\bar{\epsilon}^{[I}\dot{\psi}^{J]}-2\,\varepsilon^{IJKL}\,\epsilon_{K}\dot{\bar{\psi}}_{L}\,.\label{187_tr}
\eea

\subsubsection{Derivation from multiplet ${\bf (2,8,6)}$}\label{286}
Here, we show how invariant actions can be derived from actions of the multiplet ${\bf (2,8,6)}$. This way of getting the actions of ${\bf (1,8,7)}$ proves to be most convenient.

The irreducible multiplet ${\bf (2,8,6)}$ is described by a chiral superfield \cite{ILS2019symmetry} satisfying the constraints
\bea
    &&D^{I}\bar{\Phi} = 0,\qquad \bar{D}_{J}\Phi = 0,\qquad  D^I D^J\Phi -\frac{1}{2}\,\varepsilon^{IJKL}\,\bar{D}_K\bar{D}_L\bar{\Phi}=0,\qquad \overline{\left(\Phi\right)}=\bar{\Phi}.\label{286constr}
\eea
It has the following $\theta$-expansions over the chiral superspace $\left(t_{\rm L}\,,\theta\right)$:
\bea
    \Phi\left(t_{\rm L}\,,\theta\right) &=& \phi + \sqrt{2}\,\theta_{I}\psi^{I} -\frac{i}{\sqrt{2}}\,\theta_{I}\theta_{J}A^{IJ}
    -\frac{\sqrt{2}}{3}\,i\,\varepsilon^{IJKL}\,\theta_{I}\theta_{J}\theta_{K}\dot{\bar{\psi}}_L+4\left(\theta\right)^4\ddot{\bar{\phi}},\nn
    &&\overline{\left(\phi\right)}=\bar{\phi},\qquad \overline{\left(\psi^I\right)}=\bar{\psi}_I\,,\qquad\overline{\left(A^{IJ}\right)}=A_{IJ}=\frac{1}{2}\,\varepsilon_{IJKL}\,A^{KL}.
\eea
The component fields transform as
\bea
    &&\delta \phi = -\,\sqrt{2}\,\epsilon_{I}\psi^I,\qquad \delta \bar{\phi} = \sqrt{2}\,\bar{\epsilon}^{I}\bar{\psi}_I\,,\nn
    &&\delta \psi^I = \sqrt{2}\,i\,\bar{\epsilon}^I\dot{\phi} + i\,\epsilon_J A^{IJ},\qquad \delta \bar{\psi}_I=-\,\sqrt{2}\,i\,\epsilon_I\dot{\bar{\phi}} - i\,\bar{\epsilon}^J A_{IJ}\,,\nn
    &&\delta A^{IJ} = 4\,\bar{\epsilon}^{[I}\dot{\psi}^{J]}-2\,\varepsilon^{IJKL}\,\epsilon_{K}\dot{\bar{\psi}}_{L}\,.
\eea
It can be shown that these transformations reduce to the transformations \eqref{187_tr} by identifying  $x$ and $C$ as
\bea
    x = \frac{1}{\sqrt{2}}\left(\phi+\bar{\phi}\right),\qquad
    C = \frac{i}{\sqrt{2}}\left(\dot{\phi}-\dot{\bar{\phi}}\right).\label{286to187}
\eea
This way, we reduce the multiplet ${\bf (2,8,6)}$ to ${\bf (1,8,7)}$. The reversed procedure is known as the oxidation procedure \cite{Gates9410}.
We can also observe this duality between two multiplets at the superfield level. Indeed, the sum $\Phi+\bar{\Phi}$ satisfies the same constraints as the superfield $X$ in \eqref{187}.

The general invariant Lagrangian of the multiplet ${\bf (2,8,6)}$ can be written as
\bea
    {\cal L}=-\,\frac{1}{4}\left[\int\,d^4\theta\,{\cal K}\left(\Phi\right)+\int\,d^4\bar{\theta}\,\bar{\cal K}\left(\bar{\Phi}\right)\right],\label{286L}
\eea
where ${\cal K}\left(\Phi\right)$ is an arbitrary holomorphic function.
The component Lagrangian reads
\bea
    {\cal L}_{\bf (2,8,6)} &=& g\left(\phi,\bar{\phi}\right)\left[\dot{\phi}\dot{\bar{\phi}}+\frac{i}{2}\left(\,\psi^{I}\dot{\bar{\psi}}_{I}-\dot{\psi}^I\bar{\psi}_I\right)+\frac{1}{4}\,A^{IJ}A_{IJ}\right]-\frac{i}{2}\left[\dot{\phi}\,\partial_{\phi}g\left(\phi,\bar{\phi}\right)-\dot{\bar{\phi}}\,\partial_{\bar{\phi}}g\left(\phi,\bar{\phi}\right)\right]\psi^{I}\bar{\psi}_{I}
    \nn
    &&-\,\frac{i}{2\sqrt{2}}\left[A_{IJ}\psi^{I}\psi^{J}\,\partial_{\phi}g\left(\phi,\bar{\phi}\right)+A^{IJ}\bar{\psi}_{I}\bar{\psi}_{J}\,\partial_{\bar{\phi}}g\left(\phi,\bar{\phi}\right)\right]\nn
    &&-\,\frac{1}{24}\, g\left(\phi,\bar{\phi}\right)\left[\varepsilon_{IJKL}\,\psi^{I}\psi^{J}\psi^{K}\psi^{L}\,\partial_{\phi}\partial_{\phi} g\left(\phi,\bar{\phi}\right)
    +\varepsilon^{IJKL}\,\bar{\psi}_{I}\bar{\psi}_{J}\bar{\psi}_{K}\bar{\psi}_{L}\,\partial_{\bar{\phi}}\partial_{\bar{\phi}} g\left(\phi,\bar{\phi}\right)\right],
\eea
where $g\left(\phi,\bar{\phi}\right)$ is a special K\"ahler metric defined as
\bea
    g\left(\phi,\bar{\phi}\right)={\cal K}^{\prime\prime}\left(\phi\right)+\bar{\cal K}^{\prime\prime}\left(\bar{\phi}\right).
\eea
To gain the  Lagrangian of the multiplet ${\bf (1,8,7)}$, we impose the requirement that the target metric $g\left(\phi,\bar{\phi}\right)$ is a function of the argument $\phi+\bar{\phi}$.
There are only two solutions:
\bea
    g_{1}=1,\qquad
    g_{2}=\frac{1}{\sqrt{2}}\left(\phi+\bar{\phi}\right).
\eea
The corresponding Lagrangians read 
\bea
    {\cal L}^{1}_{\bf (2,8,6)} &=& \dot{\phi}\dot{\bar{\phi}}+\frac{i}{2}\left(\psi^{I}\dot{\bar{\psi}}_I-\dot{\psi}^{I}\bar{\psi}_I\right)+\frac{A^{IJ}A_{IJ}}{8}\,,\label{L1}\\
    {\cal L}^{2}_{\bf (2,8,6)} &=& \frac{1}{\sqrt{2}}\left(\phi+\bar{\phi}\right)\left[\dot{\phi}\dot{\bar{\phi}}+\frac{i}{2}\left(\psi^{I}\dot{\bar{\psi}}_I-\dot{\psi}^{I}\bar{\psi}_I\right)+\frac{A^{IJ}A_{IJ}}{8}\right]-\frac{i}{2\sqrt{2}}\left(\dot{\phi}-\dot{\bar{\phi}}\right)\psi^{I}\bar{\psi}_{I}\nn
    &&-\,\frac{i}{4}\left(A_{IJ}\psi^{I}\psi^{J}+A^{IJ}\bar{\psi}_{I}\bar{\psi}_{J}\right).\label{L2}\nn
\eea
Performing the redefinitions \eqref{286to187}, we obtain two possible Lagrangians of the multiplet ${\bf (1,8,7)}$:
\bea
    &&{\cal L}^{1}_{\bf (1,8,7)} = \frac{\dot{x}^2}{2}+\frac{i}{2}\left(\psi^{I}\dot{\bar{\psi}}_I-\dot{\psi}^{I}\bar{\psi}_I\right)+\frac{C^2}{2}+\frac{A^{IJ}A_{IJ}}{8}\,,\nn
    &&{\cal L}^{2}_{\bf (1,8,7)} = x\left[\frac{\dot{x}^2}{2}+\frac{i}{2}\left(\psi^{I}\dot{\bar{\psi}}_I-\dot{\psi}^{I}\bar{\psi}_I\right)+\frac{C^2}{2}+\frac{A^{IJ}A_{IJ}}{8}\right]-\frac{C}{2}\,\psi^{I}\bar{\psi}_{I}-\frac{i}{4}\left(A_{IJ}\psi^{I}\psi^{J}+A^{IJ}\bar{\psi}_{I}\bar{\psi}_{J}\right).\label{L1L2}
\eea

\subsection{Formulation in terms of ${\cal N}=4$ superfields}\label{N4}
Here, we summarize the ${\cal N}=4$ construction of the actions \eqref{L1L2} based on the harmonic superfield description of the multiplets ${\bf (1,4,3)}$ and ${\bf (0,4,4)}$ \cite{ABC, DI0706, FI1810}.

The ${\rm SU}(2)\times{\rm SU}(2)\times{\rm SU}(2)$ formulation is given via the derivatives ${\bf D}^{i\alpha}$ and $\nabla^{iA}$, where
the indices $i$, $\alpha$ and $A$ ($i,\alpha,A=1,2$) mark doublets of three commuting ${\rm SU}(2)$ groups \cite{ABC}. The odd coordinates $\theta_{I}$ and $ \bar{\theta}^I$
are split into the two sets of Grassmann coordinates $\theta^{i\alpha}$ and $\hat{\theta}^{iA}$.
We define the new derivatives in terms of the original ones \eqref{covDSU4} as
\bea
    &&{\bf D}^{11}:=D^{1},\qquad {\bf D}^{22}:=\bar{D}_{1}\,,\qquad
    {\bf D}^{12}:=D^{2},\qquad {\bf D}^{21}:=-\,\bar{D}_{2}\,,\nn
    &&\nabla^{11}:=\bar{D}_{3},\qquad \nabla^{22}:=D^{3}\,,\qquad
    \nabla^{12}:=-\,\bar{D}_{4},\qquad \nabla^{21}:=D^{4}\,.\label{SU4toSU2}
\eea
The constraints \eqref{187} can then be rewritten as
\bea
    {\bf D}^{i(\alpha}{\bf D}^{\beta)}_{i}X = 0,\qquad {\bf D}^{(i}_{\alpha}\nabla^{j)}_{A}X = 0,\qquad  \nabla^{i(A}\nabla^{B)}_{i}X = 0,\qquad
    \left({\bf D}^{(i}_{\alpha}{\bf D}^{j)\alpha}+\nabla^{(i}_{A}\nabla^{j)A}\right)X = 0.
\eea
We define ${\cal N}=4$ superfields as
\bea
    {\cal X}:=X|_{\hat{\theta}=0}\,,\qquad
    \Psi^{iA}:=\nabla^{iA}X|_{\hat{\theta}=0}\,.\label{N4SF0}
\eea
These ${\cal N}=4$ superfields satisfy
\bea
    {\bf D}^{i(\alpha}{\bf D}^{\beta)}_{i}{\cal X} = 0,\qquad
    {\bf D}^{(i}_{\alpha}\Psi^{j)A} = 0,\qquad {\bf D}^{i\alpha}=\frac{\partial}{\partial {\bf \theta}_{i\alpha}}+i\,{\bf \theta}^{i\alpha}\,\partial_t\,.
\eea
The multiplet ${\bf (1,8,7)}$ splits into the irreducible multiplets ${\bf (1,4,3)}$ and ${\bf (0,4,4)}$ described by ${\cal X}$ and $\Psi^{iA}$, respectively.
The implicit supersymmetry is realized on them through the $\eta^{iA}$-transformations
\bea
    \delta_{\eta}{\cal X} = \eta_{iA}\Psi^{iA},\qquad
    \delta_{\eta} \Psi^{iA} = \frac{1}{2}\,\eta^{A}_{j}\,{\bf D}^{i}_{\alpha}{\bf D}^{j\alpha}{\cal X}.
\eea
The Lagrangians \eqref{L1L2} correspond to the following actions:
\bea
    &&S^{1}_{\bf (1,8,7)}=-\,\frac{1}{2}\int d\zeta_{\rm (H)}\,{\cal X}^2-\frac{1}{2}\int d\zeta^{--}_{\rm (A)}\,\Psi^{+A}\Psi^{+}_{A}\,,\label{N4action1}\\
    &&S^{2}_{\bf (1,8,7)}=-\,\frac{1}{6}\int d\zeta_{\rm (H)}\,{\cal X}^3-\frac{1}{2}\int d\zeta^{--}_{\rm (A)}\,\Psi^{+A}\Psi^{+}_{A}\,{\cal X}_{\rm (A)}\,.
    \label{N4action2}
\eea
Here, both actions were constructed within the framework of the ${\cal N}=4$ harmonic superspace \cite{IL0307} (see Appendix \ref{HSS}).
The superfield $\Psi^{+A}$ satisfies both the Grassmann-analyticity and harmonic constraints:
\bea
    {\bf D}^{+\alpha}\Psi^{+A}=0,\qquad {\bf D}^{++}\Psi^{+A}=0.
\eea
The superfield ${\cal X}$ is represented via an analytic prepotential ${\cal X}_{\rm (A)}$ as \cite{DI0611}
\bea
    {\cal X}\left(\zeta\right)=\int du\,{\cal X}_{\rm (A)}\left(\zeta_{\rm (A)}\right)\,,\qquad \delta {\cal X}_{\rm (A)}\left(\zeta_{\rm (A)}\right)={\bf D}^{++}\Upsilon^{--}\left(\zeta_{\rm (A)}\right). \label{prep1}
\eea
The prepotential is analytic but otherwise unconstrained ({\it i.e.} it exhibits an arbitrary harmonic dependence, ${\bf D}^{++}{\cal X}_{\rm (A)}\neq 0$).
The multiplet ${\bf (1,4,3)}$ arises as the appropriate Wess-Zumino gauge with respect to the gauge transformation in \eqref{prep1}.

\subsection{Octonionic covariant formulation}
The multiplet ${\bf (1,8,7)}$ was also described at the component level by making use of the octonionic structure constants \cite{Toppan0511}.
Here, we present the relevant superfield constraints. 

We define ${\cal N}=8$, $d=1$ superspace in terms of real Grassmann coordinates $\theta_{\bf a}$ and $\theta$,
where the index ${\bf a}$ (${\bf a}=1,2,3,4,5,6,7$) is a vector index associated with the group ${\rm SO}(7)$.
The infinitesimal ${\cal N}=8$ transformations are realized in this superspace as
\bea
    \delta\theta_{\bf a} = \epsilon_{\bf a},\qquad \delta\theta = \epsilon,\qquad
    \delta t = i\left(\epsilon_{\bf a}\theta_{\bf a}+\epsilon\,\theta\right).
\eea
The relevant covariant derivatives are defined as
\bea
    \textrm{D}_{\bf a}=\frac{\partial}{\partial\theta_{\bf a}} - i\,\theta_{\bf a}\,\partial_t\,,\qquad \textrm{D}=\frac{\partial}{\partial\theta} - i\,\theta\,\partial_t\,\label{covDSO7}
\eea
and are related to the complex derivatives \eqref{covDSU4} as
\bea
    D^{1}=\frac{1}{\sqrt{2}}\left(\textrm{D}_{1}+i\,\textrm{D}_{2}\right),\quad D^{2}=\frac{1}{\sqrt{2}}\left(\textrm{D}_{3}-i\,\textrm{D}\right),\quad D^{3}=\frac{1}{\sqrt{2}}\left(\textrm{D}_{4}-i\,\textrm{D}_{5}\right),\quad D^{4}=\frac{1}{\sqrt{2}}\left(\textrm{D}_{6}+i\,\textrm{D}_{7}\right).\label{SU4toSO7}
\eea
The constraints \eqref{187} are then rewritten in the form
\bea
    \left(c_{\bf abc}\,\textrm{D}_{\bf c}\textrm{D}+\textrm{D}_{[{\bf a}}\,\textrm{D}_{{\bf b}]}\right)X=0,\label{187SO7}
\eea
where $c_{\bf abc}$ are the octonionic structure constants (see Appendix \ref{AppSO7}).
The octonionic covariance implies that the constraints \eqref{187SO7} are invariant under the following ${\rm SO}(7)$-transformations:
\bea
    \delta \textrm{D}_{\bf a} = \lambda_{\bf ab}\,\textrm{D}_{\bf b}+\frac{1}{2}\,c_{\bf abcd}\,\lambda_{\bf bc}\,\textrm{D}_{\bf d} - \frac{1}{2}\,c_{\bf abc}\,\lambda_{\bf bc}\,\textrm{D},\qquad
    \delta \textrm{D} = \frac{1}{2}\,c_{\bf abc}\,\lambda_{\bf ab}\,\textrm{D}_{\bf c}\,,\qquad \lambda_{\bf ab}=-\,\lambda_{\bf ba}\,.\label{covDSO7_tr}
\eea
Indeed, these transformations correspond to the nonstandard embedding of the algebra $so(7)$ in $so(8)$ (see Section 2.1.3 of \cite{KN2504}).
In fact, the total set of ${\cal N}=8$ covariant derivatives form the eight-dimensional spin representation of the group ${\rm Spin}(7)$ \cite{Varadarajan2001}
and one could think that they form a vector of ${\rm Spin}(8)$, implying the latter to be the full covariance group. Nevertheless,
the constraints \eqref{187SO7} are covariant only with respect to the ${\rm SO}(7)$ internal symmetry.

The superfield $X$ is expanded as
\bea
    X &=& x + i\,\theta_{\bf a}\psi_{\bf a} + i\,\theta\,\psi + i\,\theta\,\theta_{\bf a}A_{\bf a} + \frac{i}{2}\,\theta_{\bf a}\theta_{\bf b}\,c_{\bf abc}\,A_{\bf c}-\frac{1}{2}\,\theta\,\theta_{\bf a}\theta_{\bf b}\,c_{\bf abc}\,\dot{\psi}_{\bf c}+\frac{1}{6}\,\theta_{\bf a}\theta_{\bf b}\theta_{\bf c}\,c_{\bf abc}\,\dot{\psi}+\frac{1}{6}\,\theta_{\bf a}\theta_{\bf b}\theta_{\bf c}\,c_{\bf abcd}\,\dot{\psi}_{\bf d}\nn
    &&+\,\frac{1}{6}\,\theta\,\theta_{\bf a}\theta_{\bf b}\theta_{\bf c}\,c_{\bf abc}\,\ddot{x}-\frac{1}{24}\,\theta_{\bf a}\theta_{\bf b}\theta_{\bf c}\theta_{\bf d}\,c_{\bf abcd}\,\ddot{x}+\frac{1}{6}\,\theta_{\bf a}\theta_{\bf b}\theta_{\bf c}\theta_{\bf d}\,c_{\bf abc}\,\dot{A}_{\bf d}+\frac{1}{6}\,\theta\,\theta_{\bf a}\theta_{\bf b}\theta_{\bf c}\,c_{\bf abcd}\,\dot{A}_{\bf d}\nn
    &&+\left({\rm terms\;with\;higher}\;\theta\;{\rm powers}\right).\label{Xoct}
\eea
Taking into account \eqref{SU4toSO7}, it is straightforward to establish the direct relation between the fermionic and auxiliary components in the ${\rm SU}(4) \sim {\rm SO}(6)$ and ${\rm SO}(7)$ formulations:
\bea
    \psi_{\bf a}, \psi \longleftrightarrow \psi^I, \bar{\psi}_{I}\,,\qquad A_{\bf a} \longleftrightarrow C, A^{IJ}.
\eea
All components in \eqref{Xoct} are real and transform as
\bea
    &&\delta x = -\,i\left(\epsilon_{\bf a}\psi_{\bf a}+\epsilon\,\psi\right),\nn
    &&\delta \psi_{\bf a}=\epsilon_{\bf a}\,\dot{x}+\epsilon\,A_{\bf a}-c_{\bf abc}\,\epsilon_{\bf b}A_{\bf c}\,,\qquad \delta \psi=\epsilon\,\dot{x}-\epsilon_{\bf a}A_{\bf a}\,,\nn
    &&\delta A_{\bf a}=i\left(\epsilon_{\bf a}\dot{\psi}-\epsilon\,\dot{\psi}_{\bf a}\right)-i\,c_{\bf abc}\,\epsilon_{\bf b}\dot{\psi}_{\bf c}\,.
\eea
The Lagrangians \eqref{L1L2} are rewritten as
\bea
    &&{\cal L}^{1}_{\bf (1,8,7)} = \frac{\dot{x}^2}{2}+\frac{i}{2}\left(\psi_{\bf a}\dot{\psi}_{\bf a}+\psi\,\dot{\psi}\right)+\frac{A_{\bf a}A_{\bf a}}{2}\,,\label{L1oct}\\
    &&{\cal L}^{2}_{\bf (1,8,7)} = x\left[\frac{\dot{x}^2}{2}+\frac{i}{2}\left(\psi_{\bf a}\dot{\psi}_{\bf a}+\psi\,\dot{\psi}\right)+\frac{A_{\bf a}A_{\bf a}}{2}\right]-\frac{i}{2}\,A_{\bf a}\left(\psi\,\psi_{\bf a}+\frac{c_{\bf abc}}{2}\,\psi_{\bf b}\psi_{\bf c}\right).\label{L2oct}
\eea
They were found at the component level in \cite{Toppan0511}.
Perhaps, the harmonic superspaces associated with the generalized flag manifolds of the groups ${\rm O}(7)$ or ${\rm Spin}(7)$ \cite{Howe94}
could be used to properly rewrite the constraints on the scalar superfield $X$ and
to construct the Lagrangians \eqref{L1oct}-\eqref{L2oct} in terms of the appropriate analytic superfields. We will not dwell on these issues  here.

\subsubsection{Remark}\label{remark1}
As was emphasized, the constraints \eqref{187SO7} are covariant with respect to the symmetry ${\rm SO}(7)$.
The free Lagrangian \eqref{L1oct} is exceptional since on shell it possesses the extended symmetry ${\rm SO}(8)$.
Eliminating the auxiliary field $A_{\bf a}$\,, we obtain the relevant on-shell Lagrangian as
\bea
    {\cal L}^{1}_{\bf (1,8,7)}\big|_{\rm on-shell} = \frac{\dot{x}^2}{2}+\frac{i}{2}\left(\psi_{\bf a}\dot{\psi}_{\bf a}+\psi\,\dot{\psi}\right).\label{187Lfree}
\eea
It is invariant under the hidden ${\rm SO}(8)/{\rm SO}(7)$ transformations of the fermionic fields:
\bea
    \delta \psi = \rho_{\bf a}\psi_{\bf a}\,,\qquad \delta \psi_{\bf a} = -\,\rho_{\bf a}\psi.
\eea
The field $x$ is a trivial singlet with respect to ${\rm SO}(8)$. The Lagrangian \eqref{187Lfree} is invariant under  ${\rm SO}(8)$,
to which the ${\rm SO}(7)$ symmetry of the off-shell model is upgraded on shell.

To show exclusivity of the Lagrangian above, let us now consider the free Lagrangian \eqref{L1} of the multiplet ${\bf (2,8,6)}$.
The on-shell Lagrangian reads
\bea
    {\cal L}^{1}_{\bf (2,8,6)}\big|_{\rm on-shell} = \dot{\phi}\dot{\bar{\phi}}+\frac{i}{2}\left(\psi^{I}\dot{\bar{\psi}}_I-\dot{\psi}^{I}\bar{\psi}_I\right).\label{286Lfree}
\eea
One can show that the fermionic part is invariant under the group ${\rm SO}(8)$,
while the bosonic part is invariant under its subgroup ${\rm SO}(2)\sim {\rm U}(1)$.
Thus, the on-shell Lagrangian \eqref{286Lfree} possesses the symmetry ${\rm SO}(2)\times{\rm SO}(6)$
that just corresponds to the full covariance group of the constraints \eqref{286constr}.


\section{Indecomposable multiplet: version I}\label{v1}
In this and next Sections, we present two modified versions of the constraints \eqref{187}.
It will be  convenient to start with the version studied in \cite{FI2402}.

We consider the version\footnote{This version of ${\bf (8,8,0)}$ was denoted in \cite{ILS1807} as version 2.}
of ${\bf (8,8,0)}$ described by a complex superfield $Z^{aI}$ ($a=1,2$).
This superfield satisfies the constraints
\bea
    &&D^{(I} Z^{aJ)}=0,\qquad \bar{D}_{(I}\bar{Z}^{a}_{J)}=0,\qquad
    \overline{\left(Z^{aI}\right)}=\bar{Z}^{a}_{I}\,,\nn
    &&D^{I} Z^{aJ} = \frac{1}{2}\,\varepsilon^{IJKL}\,\bar{D}_{K}\bar{Z}^{a}_{L}\,,\qquad
    D^{J}\bar{Z}^{a}_{I} = \frac{\delta^J_I}{4}\,D^K\bar{Z}^{a}_{K}\,,\qquad \bar{D}_JZ^{aI}=\frac{\delta^I_J}{4}\,\bar{D}_KZ^{aK}.
\eea
The index $a$ ($a=1,2$) is an external vector index\footnote{The same type of indices was exploited for the semi-dynamical multiplets in \cite{FI2402}.
The accompanying ${\rm SO}(2)$ antisymmetric tensor satisfies
    $$\varepsilon^{ab}=-\,\varepsilon^{ba},\qquad \varepsilon^{12}=-\,\varepsilon^{21}=1,\qquad \overline{\left(\varepsilon^{ab}\right)}=\varepsilon^{ab},\qquad \varepsilon^{ab}\varepsilon^{cd}=\varepsilon^{ac}\varepsilon^{bd}-\varepsilon^{ad}\varepsilon^{bc}.$$}.
The field content ${\bf (8,8,0)}$ is doubled due to this vector index, {\it i.e.} the above superfields describe two irreducible multiplets ${\bf (8,8,0)}$.
This doubling is necessary for getting the correct
Wess-Zumino type term for bosonic components.
The $\theta$-expansion of $Z^{aI}$ exhibits component fields as
\bea
    Z^{aI}\left(t,\theta,\bar{\theta}\right)&=&z^{aI}-2\,\theta_J\chi^{aIJ}-\sqrt{2}\,\bar{\theta}^{I}\chi^{a} + \left({\rm terms\;with\;higher}\;\theta\;{\rm powers}\right),\nn
    &&\overline{\left(z^{aI}\right)}=\bar{z}^{a}_{I}\,,\qquad \overline{\left(\chi^a\right)}=\bar{\chi}^{a},\qquad \chi^{aIJ}=-\,\chi^{aJI}\qquad \overline{\left(\chi^{aIJ}\right)}
    =\frac{1}{2}\,\varepsilon_{IJKL}\,\chi^{aKL}=\chi^{a}_{IJ}\,.
\eea
The components transform as
\bea
    &&\delta z^{aI} = 2\,\epsilon_J\chi^{aIJ}+\sqrt{2}\,\bar{\epsilon}^{I}\chi^a,\qquad
    \delta \bar{z}^{a}_{I} = -\,2\,\bar{\epsilon}^J\chi^{a}_{IJ}-\sqrt{2}\,\epsilon_{I}\bar{\chi}^{a},\nn
    &&\delta \chi^a = -\,\sqrt{2}\,i\,\epsilon_I \dot{z}^{aI},\qquad
    \delta \bar{\chi}^a = \sqrt{2}\,i\,\bar{\epsilon}^I\dot{\bar{z}}^{a}_{I}\,,\qquad
    \delta \chi^{aIJ} = 2i\,\bar{\epsilon}^{[I}\dot{z}^{aJ]} - i\,\varepsilon^{IJKL}\,\epsilon_K\dot{\bar{z}}^{a}_{L}\,.\label{v2_tr}
\eea
Next, we consider a deformed superfield $X_{(\kappa)}$ that contains in its definition $Z^{aI}$ as an invariant subset.

The deformed superfield $X_{(\kappa)}$ is defined by the constraints
\bea
    &&D^I\bar{D}_J X_{(\kappa)} - \frac{\delta^I_J}{4}\,D^K\bar{D}_K X_{(\kappa)} = i\kappa\,\varepsilon^{ab}\left(Z^{aI}\bar{Z}^{b}_{J}-\frac{\delta^I_J}{4}\,Z^{aK}\bar{Z}^{b}_{K}\right),\nn
    &&D^I D^J X_{(\kappa)} - \frac{1}{2}\,\varepsilon^{IJKL}\,\bar{D}_K\bar{D}_L X_{(\kappa)}=i\kappa\,\varepsilon^{ab}\left(Z^{a[I}Z^{bJ]}-\frac{1}{2}\,\varepsilon^{IJKL}\,\bar{Z}^{a}_{K}\bar{Z}^{b}_{L}\right),\label{constraints_v2}
\eea
where $\kappa$ is a deformation parameter. It amounts to a reducible but indecomposable representation of supersymmetry,
and has an invariant subset formed by $Z^{aI}$. The latter appear in the expansion of $X_{(\kappa)}$\,,
\bea
    X_{(\kappa)}\left(t,\theta,\bar{\theta}\right) &=& X\left(t,\theta,\bar{\theta}\right)-\frac{i}{4}\,\kappa\,\varepsilon^{ab}\left(\theta_{I}\theta_{J}-\frac{1}{2}\,\varepsilon_{IJKL}\,\bar{\theta}^K\bar{\theta}^L\right)\left(z^{aI}z^{bJ}-\frac{1}{2}\,\varepsilon^{IJKL}\,\bar{z}^{a}_{K}\bar{z}^{b}_{L}\right)\nn
    &&-\,i\kappa\,\varepsilon^{ab}\,\bar{\theta}^J\theta_I\left(z^{aI}\bar{z}^{b}_{J}-\frac{\delta^{I}_{J}}{4}\,z^{aK}\bar{z}^{b}_{K}\right)+\kappa\left({\rm terms\;with\;higher}\;\theta\;{\rm powers}\right),
\eea
and imply the following deformed transformations for the ${\bf (1,8,7)}$ field set:
\bea
    \delta x &=& \bar{\epsilon}^{I}\bar{\psi}_{I}-\epsilon_{I}\psi^{I},\nn
    \delta \psi^{I} &=& i\,\epsilon_{J}A^{IJ}+\bar{\epsilon}^{I}\left(i\dot{x}+C\right)-i\kappa\,\varepsilon^{ab}\,\bar{\epsilon}^J\left(z^{aI}\bar{z}^{b}_{J}-\frac{\delta^I_J}{4}\,z^{aK}\bar{z}^{b}_{K}\right)+\frac{i}{2}\,\kappa\,\varepsilon^{ab}\,\epsilon_{J}\left(z^{aI}z^{bJ}-\frac{1}{2}\,\varepsilon^{IJKL}\,\bar{z}^{a}_{K}\bar{z}^{b}_{L}\right),\nn
    \delta \bar{\psi}_{I} &=& -\,i\,\bar{\epsilon}^{J}A_{IJ}-\epsilon_{I}\left(i\dot{x}-C\right)-i\kappa\,\varepsilon^{ab}\,\epsilon_J\left(z^{aJ}\bar{z}^{b}_{I}-\frac{\delta^J_I}{4}\,z^{aK}\bar{z}^{b}_{K}\right)+\frac{i}{2}\,\kappa\,\varepsilon^{ab}\,\bar{\epsilon}^{J}\left(\frac{1}{2}\,\varepsilon_{IJKL}\,z^{aK}z^{bL}-\bar{z}^{a}_{I}\bar{z}^{b}_{J}\right),\nn
    \delta C &=& -\,i\left(\bar{\epsilon}^{I}\dot{\bar{\psi}}_{I}+\epsilon_{I}\dot{\psi}^{I}\right) + \frac{3\sqrt{2}}{4}\,i\kappa \,\varepsilon^{ab}\left(\bar{\epsilon}^{I}\chi^{a}\bar{z}^{b}_{I}+\epsilon_{I}\bar{\chi}^{a}z^{bI}\right)+\frac{i}{2}\,\kappa\,\varepsilon^{ab}\left(\epsilon_{I}\chi^{aIJ}\bar{z}^{b}_{J} + \bar{\epsilon}^{I}\chi^{a}_{IJ}z^{bJ}\right),\nn
    \delta A^{IJ} &=& 4\,\bar{\epsilon}^{[I}\dot{\psi}^{J]}-2\,\varepsilon^{IJKL}\,\epsilon_{K}\dot{\bar{\psi}}_{L}-2\sqrt{2}\,\kappa\,\varepsilon^{ab}\,\bar{\epsilon}^{[I}\bar{\chi}^{a}z^{bJ]}+\sqrt{2}\,\kappa\,\varepsilon^{IJKL}\,\varepsilon^{ab}\,\epsilon_{K}\chi^{a}\bar{z}^{b}_{L}\nn
    &&-\,\sqrt{2}\,\kappa\,\varepsilon^{ab}\,\bar{\epsilon}^{[I}\chi^{a}z^{bJ]} +\frac{\sqrt{2}}{2}\,\kappa\,\varepsilon^{IJKL}\,\varepsilon^{ab}\,\epsilon_{K}\bar{\chi}^{a}\bar{z}^{b}_{L}-\kappa\,\varepsilon^{IJKL}\,\varepsilon^{ab}\,\bar{\epsilon}^M\chi^{a}_{KM}\bar{z}^{b}_{L}+2\kappa\,\varepsilon^{ab}\,\bar{\epsilon}^K\chi^{aIJ}\bar{z}^{b}_{K}\nn
    &&+\,\kappa\,\varepsilon^{ab}\,\epsilon_K\chi^{aIK}z^{bJ}-\kappa\,\varepsilon^{ab}\,\epsilon_K\chi^{aJK}z^{bI}-2\kappa\,\varepsilon^{ab}\,\epsilon_K\chi^{aIJ}z^{bK}.\label{v2_longtr}
\eea
These transformations should be complemented by \eqref{v2_tr}. The limit $\kappa=0$ leads to $X$ satisfying \eqref{187}, and the multiplet becomes fully reducible.

Now, our purpose is to construct invariant Lagrangians corresponding to this multiplet.
We deform the component Lagrangians \eqref{L1L2} and require them to be invariant under
the transformations \eqref{v2_tr} and \eqref{v2_longtr}. This ``brute force'' construction uniquely implies the deformations of both Lagrangians.
It turns out that the Lagrangian ${\cal L}^{2}_{\bf (1,8,7)}$ has no an invariant deformation. At the same time, the invariant deformation of ${\cal L}^{1}_{\bf (1,8,7)}$ exists and is written as
\bea
    {\cal L}_{\rm indec.\,I} &=& \frac{\dot{x}^2}{2}+\frac{i}{2}\left(\psi^{I}\dot{\bar{\psi}}_I-\dot{\psi}^{I}\bar{\psi}_I\right)+\frac{C^2}{2}+\frac{A^{IJ}A_{IJ}}{8}+\frac{\kappa}{8}\,\varepsilon^{ab}\left(z^{aI}z^{bJ}A_{IJ}+\bar{z}^{a}_{I}\bar{z}^{b}_{J}A^{IJ}\right)-\frac{i}{4}\,\kappa\,\varepsilon^{ab}\,C\,z^{aI}\bar{z}^{b}_{I}\nn
    &&+\,i\kappa\,\varepsilon^{ab}\left[\left(\frac{1}{\sqrt{2}}\,\bar{\chi}^{a}z^{bI}+\chi^{aIJ}\bar{z}^{b}_{J}\right)\bar{\psi}_{I}-\psi^{I}\left(\frac{1}{\sqrt{2}}\,\chi^{a}\bar{z}^{b}_{I}+\chi^{a}_{IJ}z^{bJ}\right)\right]+i\kappa\,\varepsilon^{ab}\,x\left(\chi^{a}\bar{\chi}^{b}+\frac{1}{2}\,\chi^{aIJ}\chi^{b}_{IJ}\right)\nn
    &&+\,\frac{\kappa}{2}\,\varepsilon^{ab}\,x\left(\dot{z}^{aI}\bar{z}^{b}_{I}-z^{aI}\dot{\bar{z}}^{b}_{I}\right)+\frac{3\kappa^2}{32}\,\varepsilon^{ab}\,\varepsilon^{cd}\left(2z^{aI}\bar{z}^{b}_{J}z^{cJ}\bar{z}^{d}_{I}-z^{aI}\bar{z}^{b}_{I}z^{cJ}\bar{z}^{d}_{J}\right).\label{L_v2}
\eea
It precisely coincides with the Lagrangian constructed in terms of ${\cal N}=4$ superfields in Ref. \cite{FI2402} (eq. 4.29 there).
One observes that the bosonic fields $z^{aI}$ have only the first-order time derivatives,
{\it i.e.} the pair of multiplets ${\bf (8,8,0)}$ are semi-dynamical.

\subsection{Passing on shell}
We eliminate the auxiliary fields by their equations of motion:
\bea
    &&C=\frac{i}{4}\,\kappa\,\varepsilon^{ab}\,z^{aI}\bar{z}^{b}_I,\qquad A^{IJ} = -\,\frac{\kappa}{2}\,\varepsilon^{ab}\left(z^{aI}z^{bJ}+\frac{1}{2}\,\varepsilon^{IJKL}\,\bar{z}^{a}_{K}\bar{z}^{a}_{L}\right),\nn
    &&\chi^{a}_{IJ}=-\,\frac{1}{x}\left(\bar{\psi}_{[I}\bar{z}^{a}_{J]}+\frac{1}{2}\,\varepsilon_{IJKL}\,\psi^{K}z^{aL}\right),\qquad
    \chi^{a}=-\,\frac{1}{\sqrt{2}\,x}\,\bar{\psi}_{I}z^{aI},\qquad \bar{\chi}^{a}=-\,\frac{1}{\sqrt{2}\,x}\,\psi^{I}\bar{z}^{a}_{I}\,.
\eea
Redefining the semi-dynamical fields as
\bea
    z^{aI} = \frac{1}{\sqrt{x}}\,y^{aI},\qquad \bar{z}^{a}_{I} = \frac{1}{\sqrt{x}}\,\bar{y}^{a}_{I}\,,
\eea
we obtain the following on-shell Lagrangian
\bea
    {\cal L}_{\rm on-shell}&=& \frac{\dot{x}^2}{2}+\frac{i}{2}\left(\psi^{I}\dot{\bar{\psi}}_{I}-\dot{\psi}^{I}\bar{\psi}_{I}\right)+\frac{\kappa}{2}\,\varepsilon^{ab}\left(\dot{y}^{aI}\bar{y}^{b}_{I}-y^{aI}\dot{\bar{y}}^{b}_{I}\right)\nn
    &&+\,\frac{i}{4x^2}\,\kappa\,\varepsilon^{ab}\left(\varepsilon_{IJKL}\,y^{aI}y^{bJ}\psi^{K}\psi^{L}+\varepsilon^{IJKL}\,\bar{y}^{a}_{I}\bar{y}^{b}_{J}\,\bar{\psi}_{K}\bar{\psi}_{L}\right)+\frac{i}{2x^2}\,\kappa\,\varepsilon^{ab}\left(2y^{aI}\bar{y}^{b}_{J}\,\psi^{J}\bar{\psi}_{I}-y^{aI}\bar{y}^{b}_{I}\,\psi^{J}\bar{\psi}_{J}\right)\nn
    &&+\,\frac{\kappa^2}{8x^2}\,\varepsilon^{ab}\,\varepsilon^{cd}\left(y^{aI}\bar{y}^{b}_{J}y^{cJ}\bar{y}^{d}_{I}-y^{aI}y^{bJ}\bar{y}^{c}_{I}\bar{y}^{d}_{J}\right).\label{on-shell_v2}
\eea
The relevant on-shell transformations are
\bea
    &&\delta x = \bar{\epsilon}^{I}\bar{\psi}_{I}-\epsilon_{I}\psi^{I},\nn
    &&\delta \psi^{I} = i\,\bar{\epsilon}^{I}\dot{x}-i\kappa\,\varepsilon^{ab}\left[\bar{\epsilon}^J\left(y^{aI}\bar{y}^{b}_{J}-\frac{\delta^I_J}{2}\,y^{aK}\bar{y}^{b}_{K}\right)+\frac{1}{2}\,\varepsilon^{IJKL}\,\epsilon_{J}\bar{y}^{a}_{K}\bar{y}^{b}_{L}\right]\frac{1}{x}\,,\nn
    &&\delta \bar{\psi}_{I} = -\,i\,\epsilon_{I}\dot{x}-i\kappa\,\varepsilon^{ab}\left[\epsilon_J\left(y^{aJ}\bar{y}^{b}_{I}-\frac{\delta^J_I}{2}\,y^{aK}\bar{y}^{b}_{K}\right)-\frac{1}{2}\,\varepsilon_{IJKL}\,\bar{\epsilon}^{J}y^{aK}y^{bL}\right]\frac{1}{x}\,,\nn
    &&\delta y^{aI} = \left[\frac{1}{2}\left(\bar{\epsilon}^{K}\bar{\psi}_{K}+\epsilon_{K}\psi^{K}\right)y^{aI}-\epsilon_J\psi^{I}y^{aJ}-\bar{\epsilon}^{I}\bar{\psi}_{J}y^{aJ}-\varepsilon^{IJKL}\,\epsilon_J\bar{\psi}_{K}\bar{y}^{a}_{L}\right]\frac{1}{x}\,,\nn
    &&\delta \bar{y}^{a}_{I} = -\left[\frac{1}{2}\left(\bar{\epsilon}^{K}\bar{\psi}_{K}+\epsilon_{K}\psi^{K}\right)\bar{y}^{a}_{I}-\bar{\epsilon}^J\bar{\psi}_{I}\bar{y}^{a}_{J}-\epsilon_{I}\,\psi^{J}\bar{y}^{a}_{J}-\varepsilon_{IJKL}\,\bar{\epsilon}^J\psi^{K}y^{aL}\right]\frac{1}{x}\,.\label{on-shell_tr_v2}
\eea
Defining new quantities in the adjoint ${\rm SO}(8)$ representation,
\bea
    R^{I}_{J}=i\kappa\,\varepsilon^{ab}\left(y^{aI}\bar{y}^{b}_{J}-\frac{\delta^{I}_{J}}{2}\,y^{aK}\bar{y}^{b}_{K}\right),\qquad
    R_{IJ}=-\,i\kappa\,\varepsilon^{ab}\,\bar{y}^{a}_{I}\bar{y}^{b}_{J}\,,\qquad \bar{R}^{IJ}=i\kappa\,\varepsilon^{ab}\,y^{aI}y^{bJ},\label{SO8_v2}
\eea
we can rewrite the on-shell transformations as
\bea
    \delta x &=& \bar{\epsilon}^{I}\bar{\psi}_{I}-\epsilon_{I}\psi^{I},\nn
    \delta \psi^{I} &=& i\,\bar{\epsilon}^{I}\dot{x}+\frac{1}{x}\left(\frac{1}{2}\,\varepsilon^{IJKL}\,\epsilon_{J}R_{KL}-\bar{\epsilon}^JR^{I}_{J}\right),\nn
    \delta \bar{\psi}_{I} &=& -\,i\,\epsilon_{I}\dot{x}+\frac{1}{x}\left(\frac{1}{2}\,\varepsilon_{IJKL}\,\bar{\epsilon}^{J}\bar{R}^{KL}-\epsilon_JR^{J}_{I}\right),\nn
    \delta R^{I}_{J} &=& \frac{1}{x}\left(\epsilon_J\psi^{K}R^{I}_{K}-\epsilon_K\psi^{I}R^{K}_{J}+\bar{\epsilon}^K\bar{\psi}_{J}R^{I}_{K}-\bar{\epsilon}^{I}\bar{\psi}_{K}R^{K}_{J}\right)\nn
    &&+\,\frac{1}{x}\,\epsilon_K\bar{\psi}_{L}\left(\varepsilon^{IKLM}R_{MJ}+\frac{\delta^{I}_{J}}{2}\,\varepsilon^{KLMN}\,R_{MN}\right)+\frac{1}{x}\,\bar{\epsilon}^K\psi^{L}\left(\varepsilon_{JKLM}\,\bar{R}^{IM}-\frac{\delta^{I}_{J}}{2}\,\varepsilon_{KLMN}\,\bar{R}^{MN}\right),\nn
    \delta R_{IJ} &=& -\,\frac{2}{x}\left[\frac{1}{2}\left(\bar{\epsilon}^{K}\bar{\psi}_{K}+\epsilon_{K}\psi^{K}\right)R_{IJ}+\bar{\epsilon}^K\bar{\psi}_{[I}R_{J]K}+\epsilon_{[I}\psi^{K}R_{J]K}-\varepsilon_{KLM[I}\,\bar{\epsilon}^K\psi^{L}R^{M}_{J]}\right],\nn
    \delta\bar{R}^{IJ} &=& \frac{2}{x}\left[\frac{1}{2}\left(\bar{\epsilon}^{K}\bar{\psi}_{K}+\epsilon_{K}\psi^{K}\right)\bar{R}^{IJ}+\epsilon_K\psi^{[I}\bar{R}^{J]K}+\bar{\epsilon}^{[I}\bar{\psi}_{K}\bar{R}^{J]K}-\varepsilon^{KLM[I}\,\epsilon_K\bar{\psi}_{L}R^{J]}_{M}\right].\label{on-shell_tr_SO8}
\eea
Imposing  the standard Hamiltonian constraints as
\bea
    p=\dot{x},\qquad p_{y^{aI}} - \frac{\kappa}{2}\,\varepsilon^{ab}\,\bar{y}^{b}_I \approx 0,\qquad p_{\bar{y}^{a}_{I}}-\frac{\kappa}{2}\,\varepsilon^{ab}\,y^{bI}\approx 0,\qquad p_{\psi^{I}} - \frac{i}{2}\,\bar{\psi}_I \approx 0,\qquad p_{\bar{\psi}_{I}}-\frac{i}{2}\,\psi^{I}\approx 0,
\eea
we introduce the Poisson and Dirac brackets:
\bea
    \left\lbrace x, p\right\rbrace_{\rm PB}=1,\qquad \left\lbrace y^{aI},\bar{y}^{b}_{J}\right\rbrace_{\rm DB} = \frac{1}{\kappa}\,\varepsilon^{ab}\,\delta^{I}_{J}\,,\qquad \left\lbrace \psi^{I},\bar{\psi}_{J}\right\rbrace_{\rm DB} = -\,i\,\delta^{I}_{J}\,.
\eea
Performing the canonical quantization, we find that the generators \eqref{SO8_v2} form the algebra $so(8)$:
\bea
    &&\left[R^{I}_{J},R^{K}_{L}\right]=\delta^{K}_{J}R^{I}_{L}-\delta^{I}_{L}R^{K}_{J}\,,\qquad \left[\bar{R}^{IJ},R_{KL}\right]=\delta^{I}_{K}R^{J}_{L}+\delta^{J}_{L}R^{I}_{K}-\delta^{I}_{L}R^{J}_{K}-\delta^{J}_{K}R^{I}_{L}\,,\nn
    &&\left[R^{I}_{J},\bar{R}^{KL}\right]=\delta^{K}_{J}\bar{R}^{IL}-\delta^{L}_{J}\bar{R}^{IK},\qquad \left[R^{I}_{J},R_{KL}\right]=\delta^{I}_{L}R_{JK}-\delta^{I}_{K}R_{JL}\,.\label{so8algebra}
\eea
The quantum Hamiltonian reads
\bea
    H&=&\frac{p^2}{2}-\frac{1}{x^2}\left[R^{I}_{J}\,\psi^{J}\bar{\psi}_{I}+\frac{1}{4}\left(\varepsilon_{IJKL}\,\psi^{I}\psi^{J}\bar{R}^{KL}-\varepsilon^{IJKL}\,\bar{\psi}_{I}\bar{\psi}_{J}R_{KL}\right)\right]\nn
    &&+\,\frac{1}{16x^2}\left(2\,R^{I}_{J}R^{J}_{I}+R_{IJ}\bar{R}^{IJ}+\bar{R}^{IJ}R_{IJ}\right).\label{H}
\eea
As was shown in \cite{KhKN2408}, this ${\rm SO}(8)$ symmetry is enlarged to the superconformal symmetry ${\rm OSp}(8|2)$.
It is worth pointing out that the presence of the superconformal symmetry ${\rm OSp}(8|2)$, as well as of the $R$-symmetry ${\rm SO}(8)$, is a specific feature
of the on-shell model (see also discussion in Section \ref{remark1}).

\subsection{${\cal N}=4$ superfield construction}
Here, we reformulate the modified constraints \eqref{constraints_v2} in terms of ${\cal N}=4$ superfields and
reproduce the ${\cal N}=4$ construction of the Lagrangian \eqref{L_v2}.

By analogy with the covariant derivatives \eqref{SU4toSU2}, we redefine the complex superfield $Z^{aI}$ as
\bea
    &&{\bf B}^{a11}:=Z^{a1},\qquad {\bf B}^{a22}:=\bar{Z}^{a}_{1}\,,\qquad
    {\bf B}^{a12}:=Z^{a2},\qquad {\bf B}^{a21}:=-\,\bar{Z}^{a}_{2}\,,\qquad \overline{\left({\bf B}^{ai\alpha}\right)}={\bf B}^{a}_{i\alpha}\,,\nn
    &&{\bf Z}^{a11}:=Z^{a3},\qquad {\bf Z}^{a22}:=\bar{Z}^{a}_{3}\,,\qquad
    {\bf Z}^{a12}:=Z^{a4},\qquad {\bf Z}^{a21}:=-\,\bar{Z}^{a}_{4}\,,\qquad \overline{\left({\bf Z}^{aiA}\right)}={\bf Z}^{a}_{iA}\,.
\eea
The new superfields satisfy the constraints
\bea
    &&{\bf D}^{(\alpha}_{i}{\bf B}_{j}^{a\beta)} = 0,\qquad \nabla^{(i}_{A}{\bf B}^{aj)}_{\alpha} = 0,\qquad
    \nabla^{(A}_{i}{\bf Z}_{j}^{aB)} = 0,\qquad
    {\bf D}^{(i}_{\alpha}{\bf Z}^{aj)}_{A} = 0,\nn
    &&{\bf D}^{i}_{\alpha}{\bf B}^{aj\alpha} = -\,\nabla^{i}_{A}{\bf Z}^{ajA},\qquad {\bf D}_{i}^{\alpha}{\bf Z}^{aiA} = \nabla_{i}^{A}{\bf B}^{ai\alpha}.
\eea
The constraints \eqref{constraints_v2} are then equivalently rewritten as
\bea
    &&{\bf D}^{i(\alpha}{\bf D}^{\beta)}_{i}X_{(\kappa)} = i\kappa\,\varepsilon^{ab}\,{\bf B}^{ai(\alpha}{\bf B}^{b\beta)}_{i},\qquad {\bf D}^{(i}_{\alpha}\nabla^{j)}_{A}X_{(\kappa)} = i\kappa\,\varepsilon^{ab}\,{\bf B}^{a(i}_{\alpha}{\bf Z}^{bj)}_{A},\qquad  \nabla^{i(A}\nabla^{B)}_{i}X_{(\kappa)} = i\kappa\,\varepsilon^{ab}\,{\bf Z}^{ai(A}{\bf Z}^{bB)}_{i},\nn
    &&\left({\bf D}^{(i}_{\alpha}{\bf D}^{j)\alpha}+\nabla^{(i}_{A}\nabla^{j)A}\right)X_{(\kappa)} = i\kappa\,\varepsilon^{ab}\left({\bf B}^{a(i}_{\alpha}{\bf B}^{bj)\alpha}+{\bf Z}^{a(i}_{A}{\bf Z}^{bj)A}\right).
\eea
Defining ${\cal N}=4$ superfields as
\bea
    {\cal X}_{(\kappa)}:=X_{(\kappa)}|_{\hat{\theta}=0}\,,\qquad
    \Psi^{iA}_{(\kappa)}:=\nabla^{iA}X_{(\kappa)}|_{\hat{\theta}=0}\,,\qquad
    {\cal B}^{ai\alpha}:={\bf B}^{ai\alpha}|_{\hat{\theta}=0}\,,\qquad
    {\cal Z}^{aiA}:={\bf Z}^{aiA}|_{\hat{\theta}=0}\,,\label{N4SF}
\eea
we obtain the following ${\cal N}=4$ constraints:
\bea
    {\bf D}^{i(\alpha}{\bf D}^{\beta)}_{i}{\cal X}_{(\kappa)} = i\kappa\,\varepsilon^{ab}\,{\cal B}^{ai(\alpha}{\cal B}^{b\beta)}_{i},\qquad
    {\bf D}^{(i}_{\alpha}\Psi^{j)A}_{(\kappa)} = i\kappa\,\varepsilon^{ab}\,{\cal B}^{a(i}_{\alpha}{\cal Z}^{bj)A},\qquad {\bf D}^{(\alpha}_{i}{\cal B}_{j}^{a\beta)} = 0,\qquad {\bf D}^{(i}_{\alpha}{\cal Z}^{aj)A} = 0.\label{N4constraints_v2}
\eea
The implicit supersymmetry is realized on the ${\cal N}=4$ superfields \eqref{N4SF} by the $\eta^{iA}$-transformations:
\bea
    &&\delta_{\eta}{\cal X}_{(\kappa)} = \eta_{iA}\Psi^{iA}_{(\kappa)},\qquad
    \delta_{\eta} \Psi^{iA}_{(\kappa)} = \frac{1}{2}\,\eta^{A}_{j}\,{\bf D}^{i}_{\alpha}{\bf D}^{j\alpha}{\cal X}_{(\kappa)}+\frac{i}{2}\,\kappa\,\varepsilon^{ab}\left(\eta^{i}_{B}\,{\cal Z}^{ajA}{\cal Z}^{bB}_{j}-\eta_{j}^{A}\,{\cal Z}^{ai}_{B}{\cal Z}^{bjB}-\eta_{j}^{A}\,{\cal B}^{ai}_{\alpha}{\cal B}^{bj\alpha}\right),\nn
    &&\delta_{\eta} {\cal B}^{ai\alpha} = -\,\frac{1}{2}\,\eta^{iA}\,{\bf D}^{j\alpha}{\cal Z}^{a}_{jA},\qquad
    \delta_{\eta} {\cal Z}^{aiA} = \frac{1}{2}\,\eta^{A}_{j}\,{\bf D}^{i}_{\alpha}{\cal B}^{aj\alpha}.
\eea
The superfields ${\cal X}_{(\kappa)}$ and $\Psi^{iA}_{(\kappa)}$ have a mutual invariant subset ${\cal B}^{ai\alpha}$.
Note that the harmonization of $\Psi^{iA}_{(\kappa)}$ leads to some nonanalytic superfield $\Psi^{+A}_{(\kappa)}$,
and so for the time being  we were not able to construct deformations of the analytic superpotentials in \eqref{N4action1} and \eqref{N4action2}.

The ${\cal N}=8$ invariant action of the indecomposable multiplet was constructed in terms of ${\cal N}=4$ harmonic superfields in \cite{FI2402}
by introducing a fermionic superfield ${\cal W}^{a}$ that satisfies the constraints
\bea
    {\bf D}^{i\alpha}{\cal W}^{a}={\cal B}^{ai\alpha},\qquad {\bf D}^{i(\alpha}{\bf D}^{\beta)}_{i}{\cal W}^{a}=0,\qquad \overline{\left({\cal W}^{a}\right)}=-\,{\cal W}^{a}.\label{W^a}
\eea
Making use of this superfield we redefine the superfields ${\cal X}_{(\kappa)}$ and $\Psi^{iA}_{(\kappa)}$ as
\bea
    {\cal X}_{(\kappa)}={\cal X}-\frac{i}{2}\,\kappa\,\varepsilon^{ab}\,{\cal W}^{a}{\cal W}^{b},\qquad \Psi^{iA}_{(\kappa)} = \Psi^{iA} + i\kappa\,\varepsilon^{ab}\,{\cal W}^{a}{\cal Z}^{biA},
\eea
and obtain that
\bea
    &&{\bf D}^{i(\alpha}{\bf D}^{\beta)}_{i}{\cal X} = 0,\qquad
    {\bf D}^{(i}_{\alpha}\Psi^{j)A} = 0,\qquad \delta_{\eta} {\cal X} = \eta_{iA}\Psi^{iA},\qquad
    \delta_{\eta} \Psi^{iA} = \frac{1}{2}\,\eta^{A}_{j}\,{\bf D}^{i}_{\alpha}{\bf D}^{j\alpha}{\cal X},\nn
    &&{\bf D}^{i(\alpha}{\bf D}^{\beta)}_{i}{\cal W}^{a} = 0,\qquad
    \delta_{\eta} {\cal W}^{a} = \eta_{iA}{\cal Z}^{aiA},\qquad
    {\bf D}^{(i}_{\alpha}{\cal Z}^{aj)A} = 0,\qquad \delta_{\eta} {\cal Z}^{aiA} = \frac{1}{2}\,\eta^{A}_{j}\,{\bf D}^{i}_{\alpha}{\bf D}^{j\alpha}{\cal W}^{a}.\label{irrN4}
\eea
Hence, the ${\cal N}=8$ multiplet \eqref{N4constraints_v2} decomposes into the sum of irreducible multiplets $\left({\cal X},\Psi^{iA}\right)$ and $\left({\cal W}^{a}, {\cal Z}^{aiA}\right)$,
while the dependence on $\kappa$ disappears from constraints but reappears once again in the invariant actions.

The ${\cal N}=8$ invariant action of the indecomposable multiplet was constructed in \cite{FI2402} as a deformation of \eqref{N4action1}:
\bea
    S_{\rm indec.\,I}&=&-\,\frac{1}{2}\int d\zeta_{\rm (H)}\left({\cal X}-\frac{i}{2}\,\kappa\,\varepsilon^{ab}\,{\cal W}^{a}{\cal W}^{b}\right)^2\nn
    &&+\,\frac{1}{2}\int d\zeta^{--}_{\rm (A)}\left(\Psi^{+A}\Psi^{+}_{A}+2i\kappa\,\varepsilon^{ab}\,{\cal W}_{\rm (A)}^{a}{\cal Z}^{b+A}\,\Psi^{+}_{A}+i\kappa\,\varepsilon^{ab}\,{\cal Z}^{a+A}{\cal Z}^{b+}_{A}\,{\cal X}_{\rm (A)}\right).\label{N4action}
\eea
The superfields $\Psi^{+A}$ and ${\cal Z}^{a+A}$ are analytic, while the superfields ${\cal X}$ and ${\cal W}^a$ can be represented by
their analytic prepotentials \eqref{prep1} and
\bea
    {\cal W}^a\left(\zeta\right)=\int du\,{\cal W}^a_{\rm (A)}\left(\zeta_{\rm (A)}\right),\qquad\delta{\cal W}^a_{\rm (A)}= {\bf D}^{++}\Omega^{a--}\left(\zeta_{\rm (A)}\right).\label{prep2}
\eea
The relevant component Lagrangian \eqref{L_v2} was obtained by performing the oxidation procedure \cite{Gates9410}
that restores the indecomposable multiplet at the component level.
The action \eqref{N4action2} has no similar deformation,
since we cannot simultaneously preserve invariance under the implicit symmetry \eqref{irrN4} and gauge transformations of the prepotentials \eqref{prep1} and \eqref{prep2}.

\subsubsection{Remark}
We can introduce a chiral fermionic superfields $\Pi^a$ that satisfies
\bea
    &&Z^{aI}=\frac{1}{\sqrt{2}}\,D^{I}\Pi^{a},\qquad \bar{Z}^{a}_{I}=-\,\frac{1}{\sqrt{2}}\,\bar{D}_{I}\bar{\Pi}^{a},\qquad \overline{\left(\Pi^{a}\right)}=\bar{\Pi}^{a},\nn
    &&D^{I}\bar{\Pi}^{a} = 0,\qquad \bar{D}_{J}\Pi^{a} = 0,\qquad  D^I D^J\Pi^{a} + \frac{1}{2}\,\varepsilon^{IJKL}\,\bar{D}_K\bar{D}_L\bar{\Pi}^{a}=0.
\eea
We can then redefine the scalar superfield $X_{(\kappa)}$ as
\bea
    X_{(\kappa)}=X-\frac{i}{4}\,\kappa\,\varepsilon^{ab}\left(\Pi^{a}-\bar{\Pi}^{a}\right)\left(\Pi^{b}-\bar{\Pi}^{b}\right),
\eea
where $X$ satisfies \eqref{187}.
So, the superfield $X_{(\kappa)}$ satisfying the constraints \eqref{constraints_v2} decomposes into the sum of $X$ and $\Pi^{a}$.
We can write the ${\cal N}=8$ counterpart of \eqref{W^a} by defining a new superfield $W^{a}$ as
\bea
    W^{a}:=\frac{1}{\sqrt{2}}\left(\Pi^{a}-\bar{\Pi}^{a}\right), \qquad {\cal W}^{a}=W^{a}|_{\hat{\theta}=0}\,.
\eea
Recalling Section \ref{286}, we can oxidize the multiplet ${\bf (1,8,7)}$ to ${\bf (2,8,6)}$ as
\bea
    X=\frac{1}{\sqrt{2}}\left(\Phi + \bar{\Phi}\right).
\eea
Unfortunately, we cannot split $X_{(\kappa)}$ into a sum of chiral and antichiral superfields,
which makes impossible to construct the appropriate deformation of the superfield Lagrangian \eqref{286L}.

\section{Indecomposable multiplet: version II}\label{v2}
Now we consider the second version of an indecomposable multiplet associated with another formulation of the multiplet ${\bf (8,8,0)}$ \cite{ILS1807}.
This multiplet ${\bf (8,8,0)}$ is described by a chiral superfield $Z^{a}$ and an antisymmetric tensor superfield $V^{aIJ}$. They satisfy the constraints
\bea
    &&D^{I}\bar{Z}^{a} = 0,\qquad \bar{D}_{J}Z^{a} = 0,\qquad  D^I D^JZ^{a} -\frac{1}{2}\,\varepsilon^{IJKL}\,\bar{D}_K\bar{D}_L\bar{Z}^{a}=0,\qquad \overline{\left(Z^{a}\right)}=\bar{Z}^{a},\nn
    &&D^{(I} V^{aJ)K}=0,\qquad \bar{D}_{(I}V^{a}_{J)K}=0,\qquad V^{aIJ}=-\,V^{aJI},\qquad\overline{\left(V^{aIJ}\right)}=V^{a}_{IJ}=\frac{1}{2}\,\varepsilon_{IJKL}\,V^{aKL},\nn
    &&\sqrt{2}\,D^{J} V^{a}_{IJ}=3\bar{D}_{I}\bar{Z}^a,\qquad \sqrt{2}\,\bar{D}_{J}V^{aIJ}=3D^{I}Z^{a}.\label{880_v1}
\eea
The index $a$ plays the same role as in Section \ref{v1}.
The $\theta$-expansions of both superfields exhibit component fields as
\bea
    Z^{a}\left(t_{\rm L}\,,\theta\right) &=& z^{a} + \sqrt{2}\,\theta_{I}\chi^{aI} + \sqrt{2}\,i\,\theta_{I}\theta_{J}\,\dot{v}^{aIJ}+\left({\rm terms\;with\;higher}\;\theta\;{\rm powers}\right),\nn
    V^{aIJ}\left(t,\theta , \bar{\theta}\right) &=& v^{aIJ}-\varepsilon^{IJKL}\,\theta_{K}\bar{\chi}^{a}_{L}+2\,\bar{\theta}^{[I}\chi^{aJ]}-\frac{\sqrt{2}\,i}{2}\,\varepsilon^{IJKL}\,\theta_{K}\theta_{L}\dot{\bar{z}}^{a}-\sqrt{2}\,i\,\bar{\theta}^{I}\bar{\theta}^{J}\dot{z}^{a}\nn
    &&+\,2i\,\bar{\theta}^{[I}\theta_{K}\,\dot{v}^{aJ]K}-i\,\varepsilon^{IJKL}\,\bar{\theta}^{M}\theta_{K}\,\dot{v}^{a}_{LM} + \left({\rm terms\;with\;higher}\;\theta\;{\rm powers}\right),\nn
    &&v^{aIJ}=-\,v^{aJI},\qquad \overline{\left(v^{aIJ}\right)}=v^{a}_{IJ}=\frac{1}{2}\,\varepsilon_{IJKL}\,v^{aKL},\qquad  \overline{\left(z^{a}\right)}=\bar{z}^{a},\qquad \overline{\left(\chi^{aI}\right)}=\bar{\chi}^{a}_I\,.
\eea
These components transform according to the rules
\bea
    &&\delta v^{aIJ} =  \varepsilon^{IJKL}\,\epsilon_{K}\bar{\chi}^{a}_{L}-2\,\bar{\epsilon}^{[I}\chi^{aJ]},\qquad \delta z^{a} = -\,\sqrt{2}\,\epsilon_{I}\chi^{aI},\qquad \delta \bar{z}^{a} = \sqrt{2}\,\bar{\epsilon}^{I}\bar{\chi}^{a}_{I}\,,\nn
    &&\delta \chi^{aI} = \sqrt{2}\,i\,\bar{\epsilon}^I \dot{z}^{a} - 2i\,\epsilon_J\dot{v}^{aIJ},\qquad \delta \bar{\chi}^{a}_I = -\,\sqrt{2}\,i\,\epsilon_I \dot{\bar{z}}^{a} + 2i\,\bar{\epsilon}^J\dot{v}^{a}_{IJ}\,.\label{v1_tr}
\eea
By analogy with \eqref{constraints_v2}, we define the deformed constraints,
\bea
    &&D^I\bar{D}_J X_{(\kappa)} - \frac{\delta^I_J}{4}\,D^K\bar{D}_K X_{(\kappa)} = i\kappa\,\varepsilon^{ab}\,V^{a}_{JK}V^{bIK},\nn
    &&D^I D^J X_{(\kappa)} - \frac{1}{2}\,\varepsilon^{IJKL}\,\bar{D}_K\bar{D}_L X_{(\kappa)}=\sqrt{2}\,i\kappa\,\varepsilon^{ab}\left(Z^{a}-\bar{Z}^{a}\right)V^{bIJ}.\label{constraints_v1}
\eea
The superfields $Z^{a}$ and $V^{bIJ}$ form invariant subsets of the indecomposable set  $(X_{(\kappa)}, Z, V)$\,. The solution of  \eqref{constraints_v1} can be expanded as
\bea
    X_{(\kappa)}\left(t,\theta,\bar{\theta}\right) &=& X\left(t,\theta,\bar{\theta}\right) -i\kappa\,\varepsilon^{ab}\,\bar{\theta}^J\theta_I\,v^{a}_{JK}v^{bIK}
    -\frac{\sqrt{2}}{4}\,i\kappa\,\varepsilon^{ab}\left(\theta_{I}\theta_{J}-\frac{1}{2}\,\varepsilon_{IJKL}\,\bar{\theta}^K\bar{\theta}^L\right)\left(z^{a}-\bar{z}^{a}\right)v^{bIJ}\nn
    &&+\,\kappa\left({\rm terms\;with\;higher}\;\theta\;{\rm powers}\right).
\eea
The full set of  ${\cal N}=8$ supersymmetry transformation properties of the component fields is given by \eqref{v1_tr} and by the following  transformations:
\bea
    \delta x &=& \bar{\epsilon}^{I}\bar{\psi}_{I}-\epsilon_{I}\psi^{I},\nn
    \delta \psi^{I} &=& i\,\epsilon_{J}A^{IJ}+\bar{\epsilon}^{I}\left(i\dot{x}+C\right)-i\kappa\,\varepsilon^{ab}\,\bar{\epsilon}^J\,v^{a}_{JK}v^{bIK}+\frac{\sqrt{2}}{2}\,i\kappa\,\varepsilon^{ab}\,\epsilon_{J}\left(z^{a}-\bar{z}^{a}\right)v^{bIJ},\nn
    \delta \bar{\psi}_{I} &=& -\,i\,\bar{\epsilon}^{J}A_{IJ}-\epsilon_{I}\left(i\dot{x}-C\right)-i\kappa\,\varepsilon^{ab}\,\epsilon_J\,v^{a}_{IK}v^{bJK}+\frac{\sqrt{2}}{2}\,i\kappa\,\varepsilon^{ab}\,\bar{\epsilon}^{J}\left(z^{a}-\bar{z}^{a}\right)v^{b}_{IJ}\,,\nn
    \delta C &=& -\,i\left(\bar{\epsilon}^{I}\dot{\bar{\psi}}_{I}+\epsilon_{I}\dot{\psi}^{I}\right)+i\kappa\,\varepsilon^{ab}\left(\bar{\epsilon}^{I}\chi^{aJ}v^{b}_{IJ}+\epsilon_{I}\bar{\chi}^{a}_{J}v^{bIJ}\right),\nn
    \delta A^{IJ} &=& 4\,\bar{\epsilon}^{[I}\dot{\psi}^{J]}-2\,\varepsilon^{IJKL}\,\epsilon_{K}\dot{\bar{\psi}}_{L}+2\kappa\,\varepsilon^{IJKL}\,\varepsilon^{ab}\,\bar{\epsilon}^{M}\bar{\chi}^{a}_{K}v^{b}_{LM} - 4\kappa\,\epsilon_{K}\chi^{a[I}v^{bJ]K}-\kappa\,\varepsilon^{ab}\left(\epsilon_{K}\chi^{aK}-\bar{\epsilon}^{K}\bar{\chi}^{a}_{K}\right)v^{bIJ}\nn
    &&-\,\sqrt{2}\,\kappa\,\varepsilon^{ab}\,\bar{\epsilon}^{[I}\chi^{aJ]}\left(z^{b}-\bar{z}^{b}\right)-\frac{\sqrt{2}}{2}\,\kappa\,\varepsilon^{ab}\,\varepsilon^{IJKL}\,\epsilon_{K}\bar{\chi}^{a}_{L}\left(z^{b}-\bar{z}^{b}\right).\label{v1_longtr}
\eea
Applying the same method as for the previous indecomposable multiplet, we can construct the component Lagrangian:
\bea
    {\cal L}_{\rm indec.\,II} &=& \frac{\dot{x}^2}{2}+\frac{i}{2}\left(\psi^{I}\dot{\bar{\psi}}_I-\dot{\psi}^{I}\bar{\psi}_I\right)+\frac{C^2}{2}+\frac{A^{IJ}A_{IJ}}{8}+\frac{\sqrt{2}}{8}\,\kappa\,\varepsilon^{ab}\,A^{IJ}\left(z^{a}+\bar{z}^{a}\right)v^{b}_{IJ}+\frac{i}{2}\,\kappa\,\varepsilon^{ab}\,C\,z^{a}\bar{z}^{b}\nn
    &&+\,i\kappa\,\varepsilon^{ab}\left[\frac{\sqrt{2}}{2}\left(\psi^{I}\bar{\chi}^{a}_{I}z^{b}+\bar{\psi}_{I}\chi^{aI}\bar{z}^{b}\right)-\psi^{I}\chi^{aJ}v^{b}_{IJ}-\bar{\psi}_{I}\bar{\chi}^{a}_{J}v^{bIJ}-x\,\chi^{aI}\bar{\chi}^{b}_{I}\right]\nn
    &&+\,\frac{\kappa}{2}\,\varepsilon^{ab}\,x\left(v^{a}_{IJ}\dot{v}^{bIJ}+z^{a}\dot{\bar{z}}^{b}-\dot{z}^{a}\bar{z}^{b}\right)+\frac{\kappa^2}{16}\,\varepsilon^{ab}\,\varepsilon^{cd}\left(z^{a}-\bar{z}^{a}\right)v^{bIJ}\left(z^{c}-\bar{z}^{c}\right)v^{d}_{IJ}\nn
    &&+\,\frac{\kappa^2}{8}\,\varepsilon^{ab}\,\varepsilon^{cd}\,v^{a}_{JK}v^{bIK}\,v^{c}_{IL}v^{dJL}.\label{L_v1}
\eea
This off-shell Lagrangian is not equivalent to \eqref{L_v2}.
Below we demonstrate the equivalence of these Lagrangians after elimination of auxiliary fields, {\it i.e.} on shell.

\subsection{Passing on shell}
We eliminate the auxiliary fields by their equations of motion:
\bea
    &&C=-\,\frac{i}{2}\,\kappa\,\varepsilon^{ab}\,z^{a}\bar{z}^{b},\qquad A^{IJ} = -\,\frac{\sqrt{2}}{2}\,\kappa\,\varepsilon^{ab}\left(z^{a}+\bar{z}^{a}\right)v^{b}_{IJ}\,,\nn
    &&\chi^{aI}=-\,\frac{1}{x}\left(\bar{\psi}_{J}v^{aIJ}+\frac{\sqrt{2}}{2}\,\psi^{I}z^{a}\right),\qquad
    \bar{\chi}^{a}_{I}=-\,\frac{1}{x}\left(\psi^{J}v^{a}_{IJ}+\frac{\sqrt{2}}{2}\,\bar{\psi}_{I}\bar{z}^{a}\right),
\eea
and redefine the semi-dynamical fields as
\bea
    z^{a} = \frac{1}{\sqrt{x}}\,y^{a},\qquad \bar{z}^{a} = \frac{1}{\sqrt{x}}\,\bar{y}^{a},\qquad
    v^{aIJ} = \frac{1}{\sqrt{x}}\,u^{aIJ}\,.
\eea
The on-shell Lagrangian reads
\bea
    {\cal L}_{\rm on-shell} &=& \frac{\dot{x}^2}{2}+\frac{i}{2}\left(\psi^{I}\dot{\bar{\psi}}_I-\dot{\psi}^{I}\bar{\psi}_I\right)+\frac{\kappa}{2}\,\varepsilon^{ab}\left(u^{a}_{IJ}\dot{u}^{bIJ}+y^{a}\dot{\bar{y}}^{b}-\dot{y}^{a}\bar{y}^{b}\right)\nn
    &&+\,\frac{\sqrt{2}\,i}{2x^2}\,\kappa\left(y^{a}u^{b}_{IJ}\psi^{I}\psi^{J}+\bar{y}^{a}u^{bIJ}\bar{\psi}_{I}\bar{\psi}_{J}\right)+\frac{i}{x^2}\,\kappa\left(u^{a}_{JK}u^{bIK}\psi^{J}\bar{\psi}_{I}+\frac{1}{2}\,y^{a}\bar{y}^{b}\psi^{I}\bar{\psi}_{I}\right)\nn
    &&-\,\frac{\kappa^2}{8x^2}\,\varepsilon^{ab}\,\varepsilon^{cd}\left(2y^{a}u^{bIJ}\bar{y}^{c}u^{d}_{IJ}-u^{a}_{JK}u^{bIK}u^{c}_{IL}u^{dJL}-y^{a}\bar{y}^{b}y^{c}\bar{y}^{d}\right).\label{on-shell_v1}
\eea
The on-shell transformations are
\bea
    &&\delta x = \bar{\epsilon}^{I}\bar{\psi}_{I}-\epsilon_{I}\psi^{I},\nn
    &&\delta \psi^{I} = i\,\bar{\epsilon}^{I}\dot{x}-i\kappa\,\varepsilon^{ab}\left[\bar{\epsilon}^J\left(u^{a}_{JK}u^{bIK}+\frac{\delta^{I}_{J}}{2}\,y^{a}\bar{y}^{b}\right)+\sqrt{2}\,\epsilon_{J}\bar{y}^{a}u^{bIJ}\right]\frac{1}{x}\,,\nn
    &&\delta \bar{\psi}_{I} = -\,i\,\epsilon_{I}\dot{x}-i\kappa\,\varepsilon^{ab}\left[\epsilon_J\left(u^{a}_{IK}u^{bJK}+\frac{\delta^{J}_{I}}{2}\,y^{a}\bar{y}^{b}\right)-\sqrt{2}\,\bar{\epsilon}^{J}y^{a}u^{b}_{IJ}\right]\frac{1}{x}\,,\nn
    &&\delta u^{aIJ} =  \left[\frac{1}{2}\left(\bar{\epsilon}^{K}\bar{\psi}_{K}+\epsilon_{K}\psi^{K}\right)u^{aIJ}+2\,\epsilon_{K}\psi^{[I}u^{aJ]K}+2\,\bar{\epsilon}^{[I}\bar{\psi}_{K}u^{aJ]K}+\sqrt{2}\,\bar{\epsilon}^{[I}\psi^{J]}y^{a}-\frac{\sqrt{2}}{2}\,\varepsilon^{IJKL}\,\epsilon_{K}\bar{\psi}_{L}\bar{y}^{a}\right]\frac{1}{x}\,,\nn
    &&\delta y^{a} = \left[\frac{1}{2}\left(\bar{\epsilon}^{I}\bar{\psi}_{I}+\epsilon_{I}\psi^{I}\right)y^a+\sqrt{2}\,\epsilon_{I}\bar{\psi}_{J}u^{aIJ}\right]\frac{1}{x}\,,\qquad \delta \bar{y}^{a} = -\left[\frac{1}{2}\left(\bar{\epsilon}^{I}\bar{\psi}_{I}+\epsilon_{I}\psi^{I}\right)\bar{y}^a+\sqrt{2}\,\bar{\epsilon}^{I}\psi^{J}u^{a}_{IJ}\right]\frac{1}{x}\,.\label{on-shell_tr_v1}
\eea
The Poisson and Dirac brackets are defined as follows:
\bea
    \left\lbrace x, p\right\rbrace_{\rm PB}=1,\qquad \left\lbrace u^{aIJ}, u^{b}_{KL}\right\rbrace_{\rm DB} = -\,\frac{1}{\kappa}\,\varepsilon^{ab}\,\delta^{I}_{[K}\delta^{J}_{L]}\,,\quad \left\lbrace y^{a},\bar{y}^{b}\right\rbrace_{\rm DB} = -\,\frac{1}{\kappa}\,\varepsilon^{ab}\,,\quad \left\lbrace \psi^{I},\bar{\psi}_{J}\right\rbrace_{\rm DB} = -\,i\,\delta^{I}_{J}\,.
\eea
Performing the canonical quantization, we define generators satisfying the algebra relations \eqref{so8algebra} as
\bea
    R^{I}_{J}=i\kappa\,\varepsilon^{ab}\left(u^{a}_{JK}u^{bIK}+\frac{\delta^{I}_{J}}{2}\,y^{a}\bar{y}^{b}\right),\qquad
    R_{IJ}=-\,\sqrt{2}\,i\kappa\,\varepsilon^{ab}\,\bar{y}^{a}u^{b}_{IJ}\,,\qquad \bar{R}^{IJ}=\sqrt{2}\,i\kappa\,\varepsilon^{ab}\,y^{a}u^{bIJ}.\label{SO8_v1}
\eea
Finally, we obtain the same Hamiltonian \eqref{H}.

One observes that the on-shell transformations \eqref{on-shell_tr_v1} are written via the quantities \eqref{SO8_v1} in the same form \eqref{on-shell_tr_SO8}.
This means that the on-shell Lagrangians \eqref{on-shell_v2} and \eqref{on-shell_v1} can be uniquely written in terms of the adjoint representation ``currents'' $R^{I}_{J}$\,, $R_{IJ}$ and $\bar{R}^{IJ}$.
We guess an ansatz invariant under \eqref{on-shell_tr_SO8} (up to total time derivatives):
\bea
    {\cal L}_{\rm on-shell}&=& \frac{\dot{x}^2}{2}+\frac{i}{2}\left(\psi^{I}\dot{\bar{\psi}}_{I}-\dot{\psi}^{I}\bar{\psi}_{I}\right)-\dot{R}^{I}_{J}\,{\cal A}^{J}_{I}\left(R\right)-\dot{\bar R}^{IJ}\,\bar{\cal A}_{IJ}\left(R\right)-\dot{R}_{IJ}\,{\cal A}^{IJ}\left(R\right)\nn
    &&+\,\frac{1}{x^2}\left[R^{I}_{J}\,\psi^{J}\bar{\psi}_{I}+\frac{1}{4}\left(\varepsilon_{IJKL}\,\psi^{I}\psi^{J}\bar{R}^{KL}-\varepsilon^{IJKL}\,\bar{\psi}_{I}\bar{\psi}_{J}R_{KL}\right)\right]-\frac{1}{8x^2}\left(R^{I}_{J}R^{J}_{I}+R_{IJ}\bar{R}^{IJ}\right),\nn
    \label{on-shell}
\eea
where ${\cal A}^{J}_{I}$, $\bar{\cal A}_{IJ}$ and ${\cal A}^{IJ}$ are some functions in the adjoint representation.
Calculating the explicit expressions for these functions is a technically complicated problem which will be tackled elsewhere.

\subsection{${\cal N}=4$ superfields}
In accordance with the notations of Section \ref{N4}, we make the redefinitions:
\bea
    &&{\bf Q}^{a11}:=V^{a14},\qquad {\bf Q}^{a22}:=V^{a23},\qquad
    {\bf Q}^{a12}:=V^{a24},\qquad {\bf Q}^{a21}:=V^{a13}\,,\nn
    &&{\bf V}^{a11}:=V^{a12},\qquad {\bf V}^{a22}:=V^{a34}\,,\qquad
    {\bf V}^{a12}:=\frac{1}{2\sqrt{2}}\left(Z^{a}-\bar{Z}^{a}\right),\qquad {\bf V}^{a}:=\frac{1}{\sqrt{2}}\left(Z^{a}+\bar{Z}^{a}\right),\nn
    &&\overline{\left({\bf Q}^{aA\alpha}\right)}={\bf Q}^{a}_{A\alpha}\,,\qquad {\bf V}^{aij}={\bf V}^{aji},\qquad \overline{\left({\bf V}^{aij}\right)}={\bf V}^{a}_{ij}\,,\qquad \overline{\left({\bf V}^{a}\right)}={\bf V}^{a}.
\eea
Then, the constraints \eqref{880_v1} are rewritten as
\bea
    &&{\bf D}^{(\alpha}_{i}{\bf Q}_{A}^{a\beta)} = 0,\qquad \nabla^{(A}_{i}{\bf Q}^{aB)}_{\alpha} = 0,\nn
    &&{\bf D}^{(i}_{\alpha}{\bf V}^{ajk)} = 0,\qquad {\bf D}^{i(\alpha}{\bf D}^{\beta)}_{i}{\bf V}^{a}=0,\qquad {\bf D}_{j\alpha}{\bf V}^{aij}=-\,\frac{3}{2}\,{\bf D}^{i}_{\alpha}{\bf V}^{a}=\frac{3}{4}\,\nabla^{i}_{A}{\bf Q}^{aA}_{\alpha},\nn
    &&\nabla^{(i}_{A}{\bf V}^{ajk)} = 0,\qquad \nabla^{i(A}\nabla^{B)}_{i}{\bf V}^{a}=0,\qquad
    \nabla_{jA}{\bf V}^{aij}=\frac{3}{2}\,\nabla^{i}_{A}{\bf V}^{a}=-\,\frac{3}{4}\,{\bf D}^{i}_{\alpha}{\bf Q}^{a\alpha}_{A},
\eea
and the superfield $X_{(\kappa)}$ satisfies now the constraints
\bea
    &&{\bf D}^{i(\alpha}{\bf D}^{\beta)}_{i}X_{(\kappa)} = -\,2i\kappa\,\varepsilon^{ab}\,{\bf Q}^{aA(\alpha}{\bf Q}^{b\beta)}_{A},\qquad  \nabla^{i(A}\nabla^{B)}_{i}X_{(\kappa)} = -\,2i\kappa\,\varepsilon^{ab}\,{\bf Q}^{a(A\alpha}{\bf Q}^{bB)}_{\alpha},\nn
    &&\left({\bf D}^{(i}_{\alpha}{\bf D}^{j)\alpha}+\nabla^{(i}_{A}\nabla^{j)A}\right)X_{(\kappa)} = -\,4i\kappa\,\varepsilon^{ab}\,{\bf V}^{a(i}_{k}{\bf V}^{bj)k},\qquad {\bf D}^{(i}_{\alpha}\nabla^{j)}_{A}X_{(\kappa)} = -\,2i\kappa\,\varepsilon^{ab}\,{\bf Q}^{a}_{A\alpha}{\bf V}^{bij}.
\eea
We pass to ${\cal N}=4$ superfields,
\bea
    &&{\cal X}_{(\kappa)}:=X_{(\kappa)}|_{\hat{\theta}=0}\,,\qquad
    \Psi^{iA}_{(\kappa)}:=\nabla^{iA}X_{(\kappa)}|_{\hat{\theta}=0}\,,\nn
    &&{\cal Q}^{aA\alpha}:={\bf Q}^{aA\alpha}|_{\hat{\theta}=0}\,,\qquad
    {\cal V}^{aij}:={\bf V}^{aij}|_{\hat{\theta}=0}\,,\qquad
    {\cal V}^{a}:={\bf V}^{a}|_{\hat{\theta}=0}\,,
\eea
and find the ${\cal N}=4$ form of the constraints
\bea
    &&{\bf D}^{i(\alpha}{\bf D}^{\beta)}_{i}{\cal X}_{(\kappa)} = -\,2i\kappa\,\varepsilon^{ab}\,{\cal Q}^{aA(\alpha }{\cal Q}^{b\beta)}_{A},\qquad
    {\bf D}^{(i}_{\alpha}\Psi^{j)A}_{(\kappa)} = -\,2i\kappa\,\varepsilon^{ab}\,{\cal Q}^{aA}_{\alpha}{\cal V}^{bij},\nn
    &&{\bf D}^{(\alpha}_{i}{\cal Q}_{A}^{a\beta)} = 0,\qquad {\bf D}^{(i}_{\alpha}{\cal V}^{ajk)} = 0,\qquad {\bf D}^{i(\alpha}{\bf D}^{\beta)}_{i}{\cal V}^{a}=0,\qquad {\bf D}_{j\alpha}{\cal V}^{aij}=-\,\frac{3}{2}\,{\bf D}^{i}_{\alpha}{\cal V}^{a}.\label{N4constraints_v1}
\eea
The implicit ${\cal N}=4$ supersymmetry transformations look like
\bea
    &&\delta_{\eta}{\cal X}_{(\kappa)} = \eta_{iA}\Psi^{iA}_{(\kappa)},\qquad
    \delta_{\eta} \Psi^{iA}_{(\kappa)} = \frac{1}{2}\,\eta^{A}_{j}\,{\bf D}^{i}_{\alpha}{\bf D}^{j\alpha}{\cal X}_{(\kappa)}-i\kappa\,\varepsilon^{ab}\left(\eta^{i}_{B}\,{\cal Q}^{aA\alpha}{\cal Q}^{bB}_{\alpha}+2\,\eta_{j}^{A}\,{\cal V}^{aik}{\cal V}^{bj}_{k}\right),\nn
    &&\delta_{\eta} {\cal Q}^{aA\alpha} = -\,\frac{2}{3}\,\eta^{iA}\,{\bf D}^{j\alpha}{\cal V}^{a}_{ij},\qquad
    \delta_{\eta} {\cal V}^{aij} = \frac{1}{2}\,\eta^{(i}_{A}\,{\bf D}^{j)}_{\alpha}{\cal Q}^{aA\alpha},\qquad
    \delta_{\eta} {\cal V}^{a} = -\,\frac{1}{2}\,\eta_{iA}\,{\bf D}^{i}_{\alpha}{\cal Q}^{aA\alpha}.
\eea
Recalling the action \eqref{N4action}, we suspect that there should exist ${\cal N}=4$ superfield construction of the component action \eqref{L_v1}.

\section{Conclusions}\label{con}

In this paper we have defined new indecomposable multiplets of ${\cal N}=8$ mechanics by deforming the constraints \eqref{187}
of the multiplet ${\bf (1,8,7)}$. There are two nonlinear deformations \eqref{constraints_v2} and \eqref{constraints_v1} related to different versions of the multiplet
${\bf (8,8,0)}$\footnote{These deformations are the first known examples of the nonlinear ${\cal N}=8$, $d=1$ superfield constraints.}.
In both cases we have constructed the invariant off-shell Lagrangians \eqref{L_v2} and \eqref{L_v1}, and shown that the Lagrangian \eqref{L_v2}
exactly coincides with the one constructed in \cite{FI2402}. We have shown that the relevant on-shell models are equivalently
described by the Lagrangians \eqref{on-shell} and \eqref{on-shell_tr_SO8}, where the spin variables enter through the functions in the adjoint representation of ${\rm SO}(8)$.

Among the problems for the future study we mention the manifestly ${\cal N}=4$ supersymmetric superfield construction of various off-shell component actions presented
in this paper (perhaps, using the appropriate harmonic formalism) and the proof of possible duality between the two models at the full superfield level.
It would be also interesting to consider ${\cal N}=4$ analog of
the ${\cal N}=8$ indecomposable multiplets as a nonlinear deformation of ${\cal N}=4$ multiplet $({\bf 1, 4, 3})$ by two spin ${\cal N}=4$ multiplets $({\bf 4, 4, 0})$.

The original work \cite{FIL0812} was formulated in terms of matrix superfields and aimed to derive ${\cal N}=4$ supersymmetric spin Calogero models by gauging \cite{DI0605,DI0611}. 
An intriguing question is whether we can generalize the constraints \eqref{constraints_v2} and \eqref{constraints_v1} to matrix superfields and derive from them ${\cal N}=8$ supersymmetric spin Calogero models.

\section{Acknowledgements}
The authors thank Sergey Fedoruk for useful discussions and comments.

\appendix

\section{Basics of ${\cal N}=4$, $d=1$ harmonic superspace}\label{HSS}
The ${\cal N}=4$, $d=1$ superspace $\zeta=\left\lbrace t,\theta^{i\alpha}\right\rbrace$
is defined by standard transformation properties:
\bea
    \delta \theta^{i\alpha}= \epsilon^{i\alpha}, \qquad  \delta t = -\,i\,\epsilon^{i\alpha}\theta_{i\alpha}\,,
    \qquad \overline{\left(\theta^{i\alpha}\right)}=-\,\theta_{i\alpha}\,,\qquad \overline{\left(\epsilon^{i\alpha}\right)}
    =-\,\epsilon_{i\alpha}\,.\label{SStr}
\eea
We extend it by the harmonic variables $u^{\pm}_{i}$:
\bea
    u^{+i}u^{-}_{j}-u^{-i}u^{+}_{j}=\delta^{i}_{j}.
\eea
Defining the new coordinates as
\bea
    t_{\rm (A)}=t-i\,\theta^{(i}_{\alpha}\theta^{j)\alpha}u^{+}_{i}u^{-}_{j},\qquad \theta^{\pm\alpha}=\theta^{i\alpha}u^{\pm}_{i},
\eea
we obtain the harmonic superspace $\zeta_{\rm (H)}=\left\lbrace t_{\rm (A)}\,, \theta^{\pm\alpha}, u^{\pm}_{i}\right\rbrace$.
It has an analytic harmonic subspace $\zeta_{\rm (A)} = \left\lbrace t_{\rm (A)}\,, \theta^{+\alpha}, u^{\pm}_{i}\right\rbrace$,
 which is closed under the supersymmetry transformations:
\bea
    \delta \theta^{+\alpha}= \epsilon^{+\alpha}, \qquad
    \delta t_{\rm (A)}= 2i\,\epsilon^{-\alpha}\theta^{+}_{\alpha},\qquad \delta u^{\pm}_{i}=0,\qquad
    \epsilon^{\pm\alpha} := \epsilon^{i\alpha}u^{\pm}_{i}.
\eea
The covariant derivatives are
\bea
    &&{\bf D}^{+\alpha} = \frac{\partial}{\partial \theta^{-}_{\alpha}}\,,\qquad
    {\bf D}^{-\alpha} = -\,\frac{\partial}{\partial \theta^{+}_{\alpha}}+2i\,\theta^{-\alpha}\,\frac{\partial}{\partial t_{\rm (A)}}\,,\nn
    &&{\bf D}^{++} = u^{+}_{i} \frac{\partial}{\partial u^{-}_{i}} + i\,\theta^{+\alpha}\theta^{+}_{\alpha}\,\frac{\partial}{\partial t_{\rm (A)}}
    + \theta^{+}_{\alpha}\,\frac{\partial}{\partial \theta^{-}_{\alpha}} \,,\nn
    &&{\bf D}^{--} = u^{-}_{i} \frac{\partial}{\partial u^{+}_{i}} + i\,\theta^{-\alpha}\theta^{-}_{\alpha}\,\frac{\partial}{\partial t_{\rm (A)}}
    + \theta^{-}_{\alpha}\,\frac{\partial}{\partial \theta^{+}_{\alpha}} \,,\nn
    &&{\bf D}^0 = u^{+}_{i} \frac{\partial}{\partial u^{+}_{i}} - u^{-}_{i} \frac{\partial}{\partial u^{-}_{i}} + \theta^{+}_{\alpha}\frac{\partial}{\partial \theta^{+}_{\alpha}}
- \theta^{-}_{\alpha}\frac{\partial}{\partial \theta^{-}_{\alpha}}\,.
\eea
The integration measures $d\zeta_{\rm (H)}$ and $d\zeta^{--}_{\rm (A)}$ over the full and analytic ${\cal N}=4$ harmonic superspaces are defined as
\bea
d\zeta_{\rm (H)}=\frac{1}{4}\,du\,dt_{\rm (A)}\,{\bf D}^{-}_{\alpha}{\bf D}^{-\alpha}{\bf D}^{+}_{\beta}{\bf D}^{+\beta},\qquad d\zeta^{--}_{\rm (A)}=\frac{1}{2}\,du\,dt_{\rm (A)}\,{\bf D}^{-}_{\alpha}{\bf D}^{-\alpha}.
\eea
Unconstrained harmonic superfields have an infinite number of components in their harmonic expansion.
In order to gain the finite number of fields one needs to impose harmonic constraint ${\bf D}^{++}q=0$
together with the Grassmann-analyticity condition:
\bea
    {\bf D}^{+\alpha}q\left(\zeta_{\rm (A)}\right)=0.
\eea
More details can be found in Ref. \cite{IL0307}.

\section{Octonionic structure constants}\label{AppSO7}
The third-rank antisymmetric tensor $c_{\bf abc}$ is also referred to as the octonionic tensor (or the octonionic structure constants) \cite{Toppan0511,Toppan1711,Toppan1912,KN2504}. It has the following nonzero components:
\bea
    c_{123}=c_{147}=c_{165}=c_{246}=c_{257}=c_{354}=c_{367}=1.
\eea
The fourth-rank tensor $c^{\bf abcd}$ is defined as
\bea
    &&c_{\bf abcd}=\frac{1}{6}\,\varepsilon_{\bf abcdefg}\,c_{\bf efg}\,,\qquad \varepsilon_{\bf abcdefg}\,c_{\bf defg}=24\,c_{\bf abc}\,,\nn
    &&\varepsilon_{1234567}=1,\qquad  c_{4567} = c_{2356} = c_{2437} = c_{1357} = c_{1346} = c_{1276} = c_{1245} = 1.
\eea
We list few useful identities for these tensors:
\bea
    &&c_{\bf abc}\,c_{\bf dec}=2\,\delta_{{\bf a}[{\bf d}}\,\delta_{{\bf e}]{\bf b}}-c_{\bf abde}\,,\qquad c_{\bf abcd}\,c_{\bf efcd}=8\,\delta_{{\bf a}[{\bf e}}\,\delta_{{\bf f}]{\bf b}}-2\,c_{\bf abef}\,,\qquad
    c_{\bf abc}\,c_{\bf dbc}=6\,\delta_{\bf ad}\,,\nn
    &&c_{\bf abcd}\,c_{\bf ebcd}=24\,\delta_{\bf ae}\,,\qquad c_{\bf abcd}\,c_{\bf efd}=2\left(\delta_{{\bf a}[{\bf e}}\,c_{{\bf f}]{\bf bc}}+\delta_{{\bf b}[{\bf e}}\,c_{{\bf f}]{\bf ca}}+\delta_{{\bf c}[{\bf e}}\,c_{{\bf f}]{\bf ab}}\right),\qquad c_{\bf abcd}\,c_{\bf ecd}=-\,4\,c_{\bf abe}\,.\nn
\eea


\begin{thebibliography}{99}

\bibitem{FIL0812}
S.~Fedoruk, E.~Ivanov and O.~Lechtenfeld, {\it Supersymmetric Calogero models by gauging}, {\it Phys. Rev. D} {\bf 79} (2009) 105015, \href{https://arxiv.org/abs/0812.4276}{\tt arXiv:0812.4276 [hep-th]}.

\bibitem{SS2410}
S.~Sidorov, {\it Dual superfield approach to supersymmetric mechanics with spin variables}, {\it J. Phys. A} {\bf 58} (2025) 025210, \href{https://arxiv.org/abs/2410.11618}{\tt arXiv:2410.11618 [hep-th]}.

\bibitem{Toppan1006}
M.~Gonzales, S.~Khodaee and F.~Toppan, {\it On non-minimal ${\cal N}=4$ supermultiplets in $1D$ and their associated sigma-models}, {\it J. Math. Phys.} {\bf 52} (2011) 013514, \href{https://arxiv.org/abs/1006.4678}{\tt arXiv:1006.4678 [hep-th]}.

\bibitem{Toppan1204}
M.~Gonzales, K.~Iga, S.~Khodaee and F.~Toppan, {\it Pure and entangled ${\cal N}=4$ linear supermultiplets and their one-dimensional sigma-models}, {\it J. Math. Phys.} {\bf 53} (2012) 103513, \href{https://arxiv.org/abs/1204.5506}{\tt arXiv:1204.5506 [hep-th]}.

\bibitem{Hubsch1310}
T.~H\"ubsch and G.~Katona, {\it A Q-Continuum of Off-Shell Supermultiplets}, {\it Adv. High Energy Phys.} {\bf 2016} (2016) 7350892, \href{https://arxiv.org/abs/1310.3256}{\tt arXiv:1310.3256 [hep-th]}.

\bibitem{CasimirEnergy}
B.~Assel, D.~Cassani, L.~Di Pietro, Z.~Komargodski, J.~Lorenzen and D.~Martelli, {\it The Casimir Energy in Curved Space and its Supersymmetric Counterpart}, {\it JHEP} {\bf 07} (2015) 043, \href{https://arxiv.org/abs/1503.05537}{\tt arXiv:1503.05537 [hep-th]}.

\bibitem{IS1509}
E.~Ivanov and S.~Sidorov, {\it Long multiplets in supersymmetric mechanics}, {\it Phys. Rev. D} {\bf 93} (2016) 065052, \href{https://arxiv.org/abs/1509.05561}{\tt arXiv:1509.05561 [hep-th]}.

\bibitem{FI2402}
S.~Fedoruk and E.~Ivanov, {\it ${\cal N}=8$ invariant interaction of dynamical and semi-dynamical ${\cal N}=4$ multiplets}, {\it Phys. Rev. D} {\bf 109} (2024) 085007, \href{https://arxiv.org/abs/2402.00539}{\tt arXiv:2402.00539 [hep-th]}.

\bibitem{KhKN2408}
E.~Khastyan, S.~Krivonos and A.~Nersessian, {\it Note on ${\cal N}=8$ supersymmetric mechanics with dynamical and semi-dynamical multiplets}, {\it Int. J. Mod. Phys. A} {\bf 40} (2025) 2450165, \href{https://arxiv.org/abs/2408.14958}{\tt arXiv:2408.14958 [hep-th]}.

\bibitem{ABC}
S.~Bellucci, E.~Ivanov, S.~Krivonos and O.~Lechtenfeld, {\it ABC of $N=8$, $d=1$ supermultiplets}, {\it Nucl. Phys. B} {\bf 699} (2004) 226-252, \href{https://arxiv.org/abs/hep-th/0406015}{\tt arXiv:hep-th/0406015}.

\bibitem{DI0706}
F.~Delduc and E.~Ivanov, {\it New Model of ${\cal N}=8$ Superconformal Mechanics}, {\it Phys. Lett. B} {\bf 654} (2007) 200-205, \href{https://arxiv.org/abs/0706.2472}{\tt arXiv:0706.2472 [hep-th]}.

\bibitem{FI1810}
S.~Fedoruk and E.~Ivanov, {\it Multiparticle ${\cal N}=8$ mechanics with $F(4)$ superconformal symmetry}, {\it Nucl. Phys. B} {\bf 938} (2019) 714-735, \href{https://arxiv.org/abs/1810.13366}{\tt arXiv:1810.13366 [hep-th]}.

\bibitem{Toppan0511}
Z.~Kuznetsova, M.~Rojas and F.~Toppan, {\it Classification of irreps and invariants of the $N$-extended Supersymmetric Quantum Mechanics}, {\it JHEP} {\bf 03} (2006) 098, \href{https://arxiv.org/abs/hep-th/0511274}{\tt arXiv:hep-th/0511274}.

\bibitem{ILS1807}
E.~Ivanov, O.~Lechtenfeld and S.~Sidorov, {\it Deformed ${\cal N}=8$ mechanics of ${\bf (8, 8, 0)}$ multiplets}, {\it JHEP} {\bf 08} (2018) 193, \href{https://arxiv.org/abs/1807.11804}{\tt arXiv:1807.11804 [hep-th]}.

\bibitem{ILS2019symmetry}
E.~Ivanov, O.~Lechtenfeld and S.~Sidorov, {\it Deformed $N=8$ Supersymmetric Mechanics}, {\it Symmetry} {\bf 11} (2019) 135.

\bibitem{Gates9410}
S.J.~Gates~Jr. and L.~Rana, {\it Ultramultiplets: A new representation of rigid 2D, N=8 supersymmetry}, {\it Phys. Lett. B} {\bf 342} (1995) 132, \href{https://arxiv.org/abs/hep-th/9410150}{\tt arXiv:hep-th/9410150}.

\bibitem{IL0307}
E.~Ivanov and O.~Lechtenfeld, {\it $N=4$ supersymmetric mechanics in harmonic superspace}, {\it JHEP} {\bf 09} (2003) 073, \href{https://arxiv.org/abs/hep-th/0307111}{\tt arXiv:hep-th/0307111}.

\bibitem{DI0611}
F.~Delduc and E.~Ivanov, {\it Gauging ${\cal N}=4$ supersymmetric mechanics II: (1,4,3) models from the (4,4,0) ones}, {\it Nucl. Phys. B} {\bf 770} (2007) 179-205, \href{https://arxiv.org/abs/hep-th/0611247}{\tt arXiv:hep-th/0611247}.

\bibitem{KN2504}
S.~Krivonos and A.~Nersessian, {\it Two faces of ${\cal N} = 7, 8$ superconformal mechanics}, {\it Phys. Rev. D} {\bf 111} (2025) 125025, \href{https://arxiv.org/abs/2504.13651}{\tt arXiv:2504.13651 [hep-th]}.

\bibitem{Varadarajan2001}
V.S.~Varadarajan, {\it ${\rm Spin}(7)$-subgroups of ${\rm SO}(8)$ and ${\rm Spin}(8)$}, {\it Expo. Math.} {\bf 19} (2001) 163-177.

\bibitem{Howe94}
P.S.~Howe and M.I.~Leeming, {\it Harmonic superspaces in low dimensions}, {\it Class. Quant. Grav.} {\bf 11} (1994) 2843-2852, \href{https://arxiv.org/abs/hep-th/9408062}{\tt arXiv:hep-th/9408062}.

\bibitem{DI0605}
F.~Delduc and E.~Ivanov, {\it Gauging ${\cal N}=4$ supersymmetric mechanics}, {\it Nucl. Phys. B} {\bf 753} (2006) 211-241, \href{https://arxiv.org/abs/hep-th/0605211}{\tt arXiv:hep-th/0605211}.

\bibitem{Toppan1711}
N.~Aizawa, Z.~Kuznetsova and F.~Toppan, {\it The quasi-nonassociative exceptional $F(4)$ deformed quantum oscillator}, {\it J. Math. Phys.} {\bf 59} (2018) 022101, \href{https://arxiv.org/abs/1711.02923}{\tt arXiv:1711.02923 [math-ph]}.

\bibitem{Toppan1912}
F.~Toppan, {\it The octonionically-induced ${\cal N}=7$ exceptional $G(3)$ superconformal quantum mechanics}, \href{https://arxiv.org/abs/1912.05596}{\tt arXiv:1912.05596 [hep-th]}.

\end{thebibliography}
\end{document}